\documentstyle[psfig,draft]{mn}

\def\gs{\mathrel{\raise0.35ex\hbox{$\scriptstyle >$}\kern-0.6em
\lower0.40ex\hbox{{$\scriptstyle \sim$}}}}
\def\ls{\mathrel{\raise0.35ex\hbox{$\scriptstyle <$}\kern-0.6em
\lower0.40ex\hbox{{$\scriptstyle \sim$}}}}
\def\tee{\mathrel{\raise1.2ex\hbox{$\scriptstyle -$}\kern-0.7em
\raise0.2ex\hbox{{$\scriptstyle \mid$}}}}
\title[The nature of faint submm-selected galaxies]
      {The nature of faint submillimetre-selected galaxies}

\author[Smail et al.]
       {Ian Smail,$^{\! 1}$ R.\,J.\ Ivison,$^{\! 2,3}$ A.\,W.\ Blain$^{4,5}$
	\& J.-P.\ Kneib$^{6}$
        \vspace*{1mm}\\
        $^1$ Department of Physics, University of Durham, South Road,
        Durham DH1 3LE\\
        $^2$ Astronomy Technology Centre, Royal Observatory, Blackford Hill,
        Edinburgh EH9 3HJ\\
        $^3$ Department of Physics \& Astronomy, University College London,
	Gower Street, London WC1E 6BT\\
	$^4$ Department of Astronomy, Caltech, Pasadena, CA\,91125, USA\\
        $^5$ Institute of Astronomy, Madingley Road, Cambridge CB3 0HA\\
        $^6$ Observatoire de Toulouse, 14 avenue E.\ Belin,
        31400 Toulouse, France}

\date{Accepted 2001 November 29. Received 2001 April 27; in original form 2001 January 22}

\pagerange{000--000}

\begin{document}

\maketitle

\begin{abstract}
We present the source catalogue for the SCUBA Lens Survey. We summarise
the results of extensive multi-wavelength observations of the 15
submillimetre-selected galaxies in the catalogue, from X-rays to radio. We
discuss the main observational characteristics of faint submillimetre
galaxies as a population, and consider their interpretation within the
framework of our understanding of galaxy formation and evolution.
\end{abstract}


\section{Introduction}

The highly successful far-infrared(IR) all-sky survey undertaken
by {\it IRAS} led to the identification of numerous highly obscured
star-forming and active galaxies in the local Universe, $z\ls 0.3$
(Soifer, Neugebauer \& Houck 1987).  These systems are some of the
most luminous galaxies at the present day and emit most of their
radiation in the far-IR waveband, although they contribute only 0.3
per cent of the local luminosity density (Sanders \& Mirabel 1996).

More recent work in the far-IR and submillimetre (submm) wavebands has
produced a similar revolution in our view of obscured galaxies in the
{\it distant} $z\gs 1$ Universe.  These observations have employed
the {\it COBE} and {\it ISO} satellites and the Submm Common-User
Bolometer Array (SCUBA; Holland et al.\ 1999) on the 15-m James Clerk
Maxwell Telescope\footnote{The JCMT is operated by the Joint Astronomy
Centre on behalf of the United Kingdom Particle Physics and Astronomy
Research Council (PPARC), the Netherlands Organisation for Scientific
Research, and the National Research Council of Canada.} (JCMT).  The new
observations have shown that the ultraluminous far-IR population evolves
more strongly than the equivalent optically-selected population and that,
in contrast to the local Universe, luminous obscured galaxies at high redshift
could contribute a substantial fraction of the total emitted radiation.

This conclusion is confirmed by comparing the energy density in the
optical (Bernstein et al.\ 2001) and far-IR/submm backgrounds (Puget et
al.\ 1996; Fixsen et al.\ 1998; Finkbeiner, Schlegel \& Davis 2000).
These backgrounds represent the cumulative energy emitted in these
wavebands across all epochs, mainly at redshift $z \sim 1$.  The
approximate equivalence of the energy density in the two regimes shows
that somewhere near half of the total radiation in the Universe came
from obscured energy sources, which could be either stars or AGN.  If
the majority of this emission is powered by radiation from stars with a
standard initial mass function (IMF), then approximately half of all
the stars that have formed by the present day could have formed in
highly obscured systems.  Clearly it is critically important to include
these highly-obscured sources in models of galaxy evolution if we are
to obtain a complete understanding of the formation and evolution of
galaxies.

The advent of sensitive submm imaging with SCUBA has allowed a number of
groups to undertake surveys for distant submm galaxies.  Results on the
number density of sources in blank fields as a function of 850-$\mu$m flux
density have been published by three groups: Hughes et al.\ (1998) worked
with a single deep map centred on the {\it Hubble Deep Field} (HDF),
while Barger et al.\ (1998, 1999b) employed a combination of deep/narrow
and wide/shallow observations of fields in the Lockman Hole and Hawaii
Survey Field regions; finally, there has been a survey of areas from the
Canada-France Redshift Survey (Eales et al.\ 1999, 2000; Lilly et al.\
1999). Shallower, wider surveys have also been carried out by Borys et
al.\ (2001) and by the UK Submm Survey consortium (Dunlop 2001; Scott et
al.\ 2001; Fox et al.\ 2001; Almaini et al.\ 2001).  Due to the modest
resolution of SCUBA, 15$''$ FWHM at 850\,$\mu$m, the deepest of these
studies are confusion limited at $\sim 2$\,mJy (Hughes et al.\ 1998);
this is the deepest flux density for which reliable source detection is
possible in blank fields (Blain, Ivison \& Smail 1998; Hogg 2001).

At slightly longer wavelengths, observations using the MAMBO 1.2-mm
camera on the IRAM 30-m telescope have recently been reported in both
field and lensing cluster regions (Bertoldi et al.\ 2000).

Rather than mapping large fields, another approach to construct large samples
of submm-detected galaxies is to exploit the tight correlation between the
far-infrared and radio luminosities of star-forming galaxies (Condon 1992)
and use SCUBA to target samples of faint radio sources (Chapman et al.\
2001b, 2001c).  This innovative technique is particularly well-suited
for identifying the wide-field samples needed to tackle issues such as the
clustering strength of SCUBA galaxies (e.g.\ Almaini et al.\ 2001).

Our collaboration has adopted yet another approach with the aim of
pushing below the confusion limit of the blank-field surveys.  We achieve
this by using massive gravitational cluster lenses to increase both the
sensitivity and resolution of SCUBA (Blain 1998).  The first submm counts
were based on maps of two clusters (Smail, Ivison \& Blain 1997). The
survey was subsequently expanded to cover seven lensing clusters at
$z=0.19$--0.41 (Smail et al.\ 1998). The results of similar surveys have
recently been reported by Chapman et al.\ (2001) and van der Werf et al.\
(2001b). These observations of lensing clusters benefit from a typical
amplification factor of 2--3$\times$, improving both the sensitivity of
the maps and their effective resolution, and so allowing confusion-free
counts to be derived down to $\sim 0.5$\,mJy (Blain et al.\ 1999a),
well below the conventional 2-mJy field confusion limit for SCUBA. In
fortuitous cases, the amplification can exceed $10\times$ (e.g.\ van der
Werf et al.\ 2001b), providing the opportunity to identify submm
galaxies as faint as $\sim 0.1$\,mJy and study their properties.

Based upon this survey we have published the number counts of submm
galaxies (Smail, Ivison \& Blain 1997; Blain et al.\ 1999a), the
identification of the counterparts to the submm sources in the optical
(Smail et al.\ 1998), near-infrared (Smail et al.\ 1999a; Frayer et al.\
2000), radio (Smail et al.\ 2000) and X-ray (Fabian et al.\ 2000) bands,
as well as optical spectroscopy of candidate counterparts (Barger et
al.\ 1999a).  We have also provided detailed multi-wavelength follow-up
observations of the brighter sources (Ivison et al.\ 1998a, 2000a).
Building on the redshifts determined for the three brightest sources in
the sample, we have obtained the first CO detections of submm-selected
galaxies using the OVRO and IRAM interferometers (Frayer et al.\
1998, 1999; Kneib et al.\ 2001). Yet higher-resolution CO images of
one source have been obtained by combining data from the OVRO and BIMA
arrays (Ivison et al.\ 2001). High-resolution mm-continuum observations
using the OVRO array have also been presented (Frayer et al.\ 2000).
Similar mm-continuum observations of sources in other SCUBA surveys have
been presented by Downes et al.\ (1999), Gear et al.\ (2000) and Lutz
et al.\ (2001).  Finally, the interpretation of these observations and
their relevance to our understanding of galaxy formation and evolution
at high redshifts has been discussed by Blain et al.\ (1999b, 1999c).
This paper includes a summary and update of these previous results.

As a benchmark for the following discussion we note that an
Ultraluminous Infrared Galaxy (ULIRG) with a far-IR luminosity of
$L_{\rm FIR} \sim 3\times 10^{12} L_\odot$ (similar to Arp\,220), and
thus a star-formation rate (SFR) of $\sim 300$\,M$_\odot$\,yr$^{-1}$
would have a 850-$\mu$m flux density of $\gs 3$\,mJy out to $z\sim
10$ in a Universe with $q_0=0.5$.\footnote{We assume $q_o=0.5$ and
$H_o=50$\,km\,s$^{-1}$\,Mpc$^{-1}$ throughout the paper, except in
Fig.~1.} The effects of different SEDs and world models on the results
in the two SCUBA observing bands are illustrated in Fig.~1.  In three
8-hr shifts of observing in good conditions with SCUBA it is possible
to achieve a 3$\sigma$ flux limit of 3\,mJy at 850\,$\mu$m across a
160$''$-diameter field (upgrades to SCUBA now mean that this limit is
reached in closer to two shifts), probing a volume of 10$^6$ Mpc$^3$
out to $z\sim 10$ for ULIRGs.

%
%
\begin{figure}
\centerline{\psfig{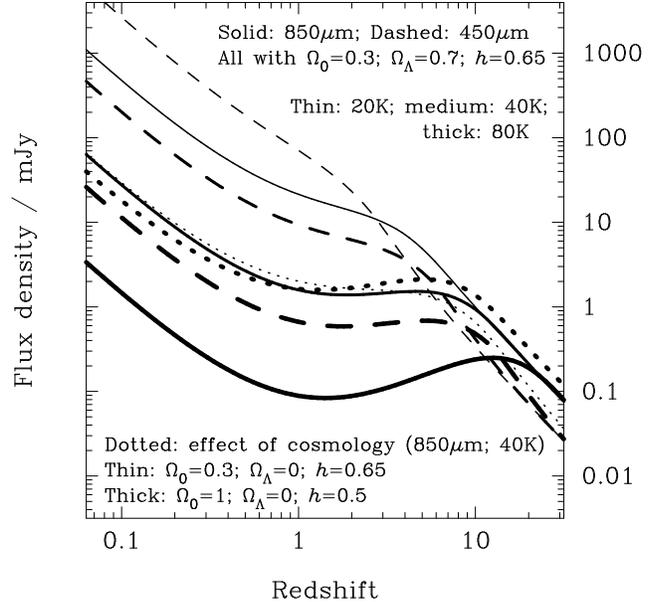} }
\noindent{\small\addtolength{\baselineskip}{-3pt}}
\caption{The flux density at 450 and 850\,$\mu$m
expected from a ULIRG-like galaxy as a function of
redshift for different dust temperatures and cosmologies.  The galaxy
is assumed to have $L_{\rm FIR} = 3 \times 10^{12}$\,L$_\odot$.}
\end{figure}

Here we present the source catalogue of luminous submm galaxies from
our SCUBA survey, and discuss the current identifications and
properties of their counterparts in other wavebands.  We then summarise
the broad characteristics of the populations that contribute to the
submm background radiation. In \S2 we briefly describe the submm
observations and their reduction and analysis. In \S3 we present our
updated estimate of the source counts at 850 and 450\,$\mu$m, and in \S4 we
describe the identification of counterparts and constraints on their
likely redshifts and power sources. A brief summary of the characteristics
of each source is given in \S5, and we discuss the properties
of the different classes of submm galaxy in \S6. In \S7 we present
the main conclusions arising from the SCUBA lens survey and
summarise these in \S8.

\section{Submm observations, reduction and analysis}

These data were obtained using SCUBA (Holland et al.\ 1999) on the
JCMT. SCUBA contains a range of detectors and detector arrays cooled
to 0.1\,{\sc k} covering the atmospheric windows from 350\,$\mu$m to
2000\,$\mu$m.  In our survey, we operated the 91-element short-wave
array at 450\,$\mu$m and the 37-element long-wave array at
850\,$\mu$m, giving half-power beam widths of $7.5''$ and $14.7''$
respectively. Operating in its conventional `stare' or `jiggle' mode
and discarding the noisy map edges, the short- and long-wave arrays
have effective fields of view of $2.3'$ and $2.8'$ respectively.
The design of the optics ensures that a suitable jiggle pattern for
the chopping secondary mirror can fully sample the image plane
simultaneously at 450 and 850\,$\mu$m.  The multiplexing and high
efficiency of the arrays means that SCUBA maps three orders of
magnitude faster than previous submm detectors.

The clusters exploited in our survey are all well-studied, massive
systems, which exhibit strongly-lensed (i.e.\ high-amplification) images
of background galaxies (most with measured spectroscopic redshifts)
that are employed to accurately model the mass distribution within the
cluster  (see Kneib et al.\ 1996; Smith et al.\ 2001a).  These models can
then be used to `correct' the submm observations to derive the true fluxes
of sources identified in our SCUBA maps (Blain et al.\ 1999a; \S3). The
clusters and field centres for our observations are listed in Table~1.

The submm maps of the seven clusters used in our analysis were obtained
from SCUBA observations on the nights of 1997 July 02--05, 1997 August
09--14 and 21--22, 1997 September 22, 1997 December 19, 1998 January
26, 29 and 31, 1998 February 01, 1998 March 12--13 and 1998 April 03.
The typical integration time was 25--35\,ks and the maps have effective
noise levels of $\sim 1.7$\,mJy rms at 850\,$\mu$m (Table~1).

The observations employed a 64-point jiggle pattern, fully sampling
both arrays over a period of 128\,s. The pattern was subdivided into
nod cycles so that the target position was be switched between the
signal and reference positions (separated by 60$''$ in azimuth) in a
repeating signal--reference--reference--signal scheme, with sixteen
1-s jiggles in each position. As well as tracing out the jiggle
pattern, the secondary was chopped at $\simeq$7\,Hz, also by $60''$ in
azimuth.

A side effect of this complex procedure is the creation of two {\it
negative} images of any source in the field, each $-1/2 \times$ the
intensity of the source and separated from the source in azimuth by
60$''$. Since SCUBA has no image rotator, the positions of these
negative sources migrate slowly around the source during each
integration. Azimuthal chopping has advantages; sky removal is
optimised and the field rotation means that additional real sources
are less likely to be consistently nullified, however it carries a
penalty: unlike nodding and chopping in R.A./Dec.\ coordinates (the mode
now used as standard; see Ivison et al.\ 2000b), information contained
in the negative images of each source is difficult to recover.
Moreover, the noise levels we quote inevitably include a contribution
from the negative images of real, bright sources as well as from faint
sources below our detection threshold (i.e.\ conventional confusion
noise).

The pointing stability was checked every hour and regular skydips were
performed to measure the atmospheric opacity, which averaged 0.2 at
850\,$\mu$m and 1.0 at 450\,$\mu$m, and was as good as 0.09 and 0.4 on
respectively on occasion. The rms pointing errors were typically $2''$
(see \S4.3).

The dedicated SCUBA data reduction software ({\sc surf}, Jenness 1997)
was used to reduce the observations. The reduction consisted of
subtracting the reference from the signal on a second-by-second basis,
giving 1280 data points per bolometer per 20-integration scan, where an
integration represents the time needed to complete a full jiggle
pattern (64\,s in both the reference and signal positions). The data
were flatfielded and corrected for atmospheric attenuation, then
inspected statistically and visually. Some were rejected on the basis
of large deviations from a bolometer's mean, some others on the basis
of noticably peculiar behaviour. Data from bolometers suffering
excessively from $1/f$ noise were also flagged as bad and removed from
the analysis.  At this stage, median values evaluated across the whole
array were used to compensate for the spatially-correlated sky
emission, reducing the effective noise-equivalent flux density to
75--90\,mJy\,Hz$^{-1/2}$ at 850\,$\mu$m (see Ivison et al.\ 1998b).

The sky position appropriate for each data point is known, based on the
flatfield (a file containing the relative bolometer sensitivities and
positions) and the jiggle pattern, so maps can be generated by placing
data from each bolometer on to an astrometric grid of $2''$ and $4''$
pixels at 450 and 850\,$\mu$m respectively (approximately Nyquist
sampling). Data values more than 4$\sigma$ from each pixel's mean were
rejected and pixels closer than 14$''$ (one beamwidth) to the most
extreme jiggle positions (regions with low effective exposure times)
were blanked out from the map.

The resulting images were calibrated using nightly beam maps of
Uranus, Mars and occasionally CRL\,618. Absolute flux calibration is
accurate to 10 and 20 per cent at 850 and 450\,$\mu$m, values fixed in
part by the 5-per-cent uncertainty in the brightness temperature of Mars.
Several calibration methods were employed and used in the
appropriate circumstances: the conventional `Jy beam$^{-1}$' method
was used primarily to calculate the noise in each map; an `aperture
photometry' method, with a variety of apertures, was used to measure
source flux densities in map regions not affected significantly by
confusion. The maps are presented in Fig.~2.

%
%
\begin{figure*}
\centerline{\psfig{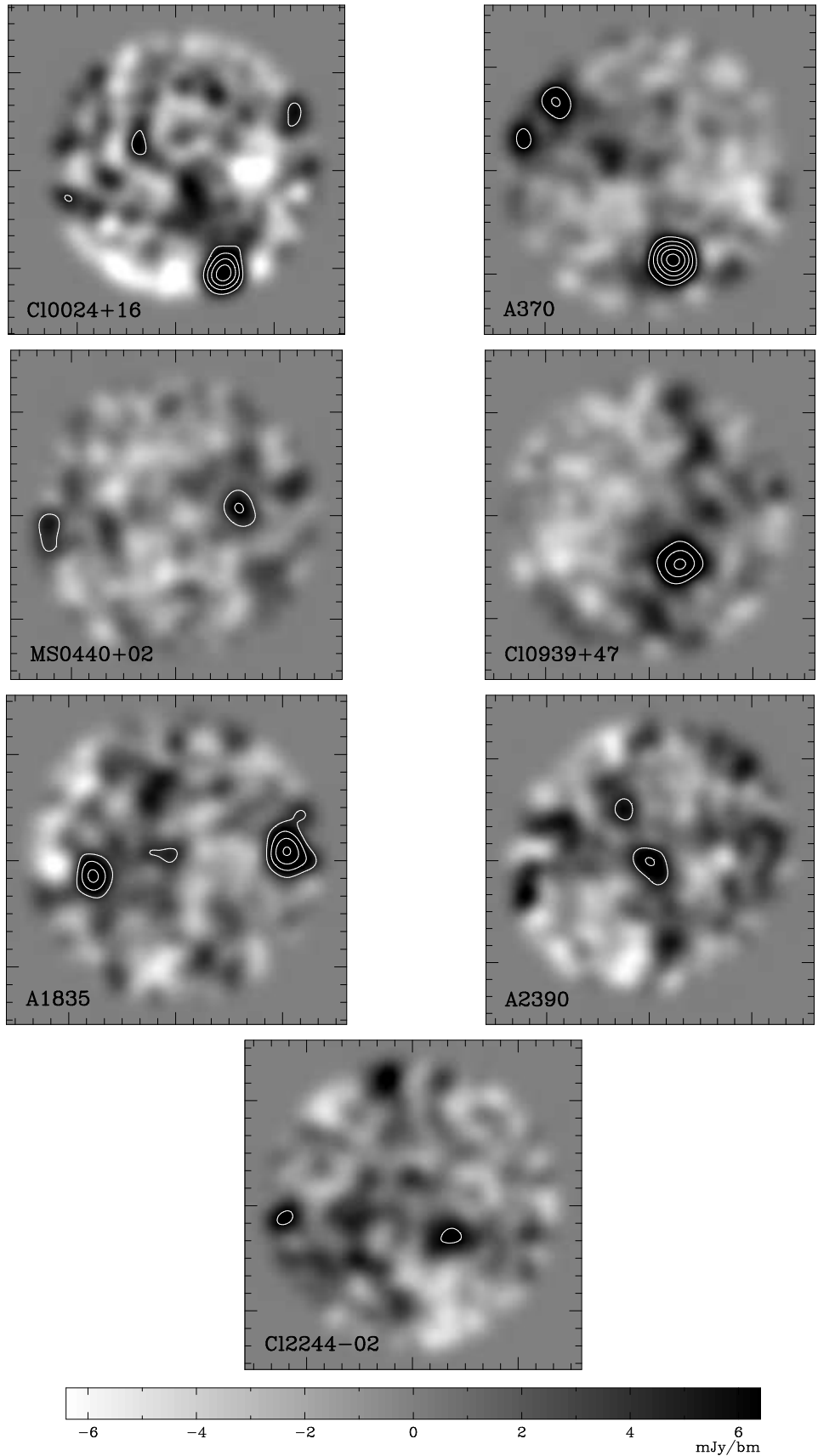} }
\caption{850-$\mu$m maps of the seven cluster fields from the SCUBA
Lens Survey. These have been convolved with a 12-$''$ FWHM Gaussian
for display purposes and the contours identify the sources in our
catalogue.  The major tickmarks are every 60$''$ from the field centre
positions given in Table~1. North is up; east to the left. Contours
are plotted at $3,4... \times \sigma$.}
\end{figure*}

Source catalogues from our fields were constructed using the
Sextractor package (Bertin \& Arnouts 1996). The detection algorithm
requires that the surface brightness in four contiguous pixels (or 64
sq.\ arcsec) exceeds a threshold after subtracting a smooth background
signal and convolving the map with a $16''\times 16''$ top-hat filter.
The detection threshold used was $1\sigma$ of the map noise (Table~1),
which includes both contributions from sources and their
negative reference beams. The `true' map noise is typically
20--30 per cent lower than the values listed in Table~1.

Our observations were obtained early in SCUBA's operation and we adopted
a conservative azimuthal chop, since it was then unclear whether
differential spillover of the beam sidelobes onto the ground could
lead to systematic errors or increased noise for R.A./Dec.-chopped
observations. This is now known not to be a significant effect. The
smearing of the negative images of sources in the final map, when using
azimuthal chopping, reduces their usefulness for {\sc clean}-type
analyses (Hughes et al.\ 1998; Ivison et al.\ 2000b) and our source
detection therefore does not use any information from the negative
images of sources. Once detected, however, the sources were checked
visually: all of the brightest sources have negative counterparts at
the expected positions (e.g.\ for the A\,370 map in Fig.~2 these lie 60$''$
to the north-west/south-east of each source). The number of independent
resolution elements in our survey is about 760, and so no 4-$\sigma$ noise
peaks are expected in the catalogue. As a check, we ran the detection
algorithm on the negative copy of each map and found only the negative
reference-beam images of the brightest sources. We were also able to
recover all of the sources from independent sub-samples of the data.

As we have not `cleaned' the maps, there is an additional source of
confusion for faint sources in those maps which contain at least one
bright source: the reference beams from bright sources could be chopped
onto another source, resulting in deviations in its apparent position
and flux density.  However, the only clear case of this is for
SMM\,J14010+0252 (the central galaxy in A\,1835), where the chopped
images of two bright sources overlap near the central galaxy, both
reducing its flux density and confusing its position (Ivison et
al.\ 2000a).

With the lensing amplification taken into account, the data shown in
Fig.~2 are some of the deepest sub-mm maps so far published. In regions
close to critical lines in the image plane the maps are sensitive to
galaxies with intrinsic flux densities as faint as 0.2\,mJy, if these
have a surface density great enough to populate the relevant regions of
the source plane.

A total of 17 sources are detected above the nominal 3-$\sigma$ limits,
and 10 above the 4-$\sigma$ limits, of the 850-$\mu$m maps (which
correspond to $\sim 4$- and $\sim 5$-$\sigma$ when using noise
estimates from cleaner regions of the maps). We list these in order of
their fluxes in Table~2.  These are detected in a total surveyed area
of $\simeq 40$\,arcmin$^2$ in the image plane (Fig.~3).  None of the sources
are resolved at 850\,$\mu$m (c.f.\ Ivison et al.\ 2001a). Five of these
sources are also detected at 450\,$\mu$m (Table~2). From the surface density of
sources detected in a typical field, we can be confident that our
observations are not limited by confusion: there are $\sim 45$ beams
per source (Blain, Ivison \& Smail 1998; Hogg 2001).

The atmospheric transparency at 450\,$\mu$m on Mauna Kea is lower than
at 850\,$\mu$m even in good weather, and together with the reduced
aperture efficiency this explains the smaller number of sources
detected in our 450-$\mu$m maps (Table~2).  The flux density limits for
detection in the 450-$\mu$m maps are listed in Table~1.

%
%
{\small
\setcounter{table}{0}
\begin{table*}
\caption{ Summary of the Observations }
\begin{tabular}{lccccccccc}
\noalign{\smallskip}\hline
\noalign{\smallskip}
Cluster & $\alpha$,$\delta$~(J2000)   & $z_{cl}$ & T$_{\rm SCUBA}$ & $\!\!\!\sigma({850}\mu{\rm m})\!\!\!$ & $\!\!\!\sigma({450}\mu{\rm m})\!\!\!$ & T$_{\rm VLA}$ & $\!\!\!\sigma({1.4}{\rm GHz})\!\!\!$ & T$_{\rm ROSAT}$ & $\!\!\!\sigma(1.0{\rm KeV})^a\!\!\!$ \cr
        &    &       &  (ks) & (mJy) & (mJy) & (ks) & ($\mu$Jy) & (ks) &  \cr
\noalign{\smallskip}\hline
\noalign{\smallskip}
Cl\,0024+16   & 00\,26\,35.80 $+$17\,09\,41.0  & 0.39  &  15.6 & 1.5 & 20 & 75.7 &  15 & 82.6 & ~2.6 \cr
A\,370        & 02\,39\,53.00 $-$01\,35\,06.0  & 0.37  &  33.8 & 1.9 & 10 & 65.8$^b$ &  10 & 29.7 & ~2.9 \cr
MS\,0440+02   & 04\,43\,09.00 $+$02\,10\,19.9  & 0.19  &  35.8 & 1.5 & 20 & 27.9 &  15 & 27.3 & ~4.2 \cr
Cl\,0939+47   & 09\,42\,56.38 $+$46\,59\,10.4  & 0.41  &  30.1 & 1.9 & ~7 & 60.0$^b$ &  ~9 & 45.8 & ~2.7 \cr
A\,1835       & 14\,01\,02.20 $+$02\,52\,43.0  & 0.25  &  23.0 & 1.7 & ~7 & 26.6$^b$ &  16 & ~2.8 & 16~~ \cr
A\,2390       & 21\,53\,36.89 $+$17\,41\,45.8  & 0.23  &  33.7 & 2.2 & 20 & ~9.7  & 100 & 32.8 & ~4.0 \cr
Cl\,2244$-$02 & 22\,47\,11.90 $-$02\,05\,38.0  & 0.33  &  25.6 & 1.7 & 20 & 27.8 &  17 & 33.7 & ~3.5 \cr
\noalign{\smallskip}\hline
\noalign{\smallskip}
\end{tabular}

$a$) 1-$\sigma$ flux limits in the 0.1--2.0\,keV band from archival
{\it ROSAT} {\it HRI} observations, in units of 10$^{-15}$ erg
s$^{-1}$ cm$^{-2}$ -- \\
$b$) excluding A-configuration data (Owen et al., in prep; Ivison et
al., 2001b).
\end{table*}
}

%
%
{\small
\setcounter{table}{1}
\begin{table*}
\begin{center}
\caption{\hfil Catalogue of Source Positions and Submm Fluxes \hfil}
\begin{tabular}{lcccccl}
\noalign{\smallskip}\hline
\noalign{\smallskip}
{Source} & {$S_{850}$} &  {$S_{450}$} & Submm  & Radio & Opt/NIR &
{ Comments ~\hfill } \cr
{} & {(mJy)} & {(mJy)} & {$\alpha$,$\delta$~(J2000)} & {$\alpha$,$\delta$~(J2000)} &
{$\alpha$,$\delta$~(J2000)} & {} \cr
\noalign{\smallskip}\hline
\noalign{\smallskip}
SMM\,J02399$-$0136 & 23.0  &  ~85   & 02\,39\,51.9 $-$01\,35\,59 & 02\,39\,52.00 $-$01\,35\,57.9 & 02\,39\,51.88 $-$01\,35\,58.0 & L1/L2$^a$ \cr
SMM\,J00266+1708   & 18.6  &  ... & 00\,26\,34.1 $+$17\,08\,32 & 00\,26\,34.06 $+$17\,08\,33.1 & 00\,26\,34.11 $+$17\,08\,33.2 & M11$^{b,g}$ \cr
SMM\,J09429+4658   & 17.2  &  ~61   & 09\,42\,54.7 $+$46\,58\,44 & 09\,42\,54.51 $+$46\,58\,44.7 & 09\,42\,54.65 $+$46\,58\,44.7 & H5$^c$ \cr
SMM\,J14009+0252   & 14.5  &  ~33   & 14\,00\,57.7 $+$02\,52\,50 & 14\,00\,57.55 $+$02\,52\,48.6 & 14\,00\,57.57 $+$02\,52\,49.1 & J5$^d$ \cr
SMM\,J14011+0252   & 12.3  &  ~42   & 14\,01\,05.0 $+$02\,52\,25 & 14\,01\,04.96 $+$02\,52\,23.5 & 14\,01\,04.97 $+$02\,52\,24.6 & J1/J2$^d$ \cr
SMM\,J02399$-$0134 & 11.0  &  ~42 & 02\,39\,56.4 $-$01\,34\,27 & 02\,39\,56.30 $-$01\,34\,30.9 & 02\,39\,56.51 $-$01\,34\,27.1 & L3$^e$ \cr
SMM\,J22471$-$0206 & ~9.2  &  $<60$ & 22\,47\,10.4 $-$02\,05\,59 & ... & 22\,47\,10.10 $-$02\,05\,57.2 & P4? \cr
SMM\,J02400$-$0134 & ~7.6  &  $<50$ & 02\,39\,57.9 $-$01\,34\,45 & ... & ... & Blank \cr
SMM\,J04431+0210   & ~7.2  &  $<60$ & 04\,43\,07.2 $+$02\,10\,24 & ... & 04\,43\,07.10 $+$02\,10\,25.1 & N4$^c$ \cr
SMM\,J21536+1742   & ~6.7  &  $<60$ & 21\,53\,38.2 $+$17\,42\,13 & ... & 21\,53\,38.31 $+$17\,42\,13.3 & K2? \cr
SMM\,J00265+1710   & ~6.1  &  $<60$ & 00\,26\,31.3 $+$17\,10\,04 & 00\,26\,31.28 $+$17\,10\,09.6 & ... & Blank? \cr
SMM\,J22472$-$0206 & ~6.1  &  $<60$ & 22\,47\,13.9 $-$02\,06\,11 & ... & 22\,47\,13.76 $-$02\,06\,07.9 & P2? \cr
SMM\,J00266+1710   & ~5.9  &  $<60$ & 00\,26\,37.9 $+$17\,09\,51 & ... & ... & Blank? \cr
SMM\,J00267+1709   & ~5.0  &  $<60$ & 00\,26\,39.7 $+$17\,09\,12 & ... & ... & Blank \cr
SMM\,J04433+0210   & ~4.5  &  $<60$ & 04\,43\,15.0 $+$02\,10\,02 & 04\,43\,15.04 $+$02\,10\,01.3 & 04\,43\,15.06 $+$02\,10\,02.8 & N5 \cr
\noalign{\medskip}
\multispan7{Cluster Galaxies \hfil} \cr
\noalign{\smallskip}
SMM\,J21536+1741   & ~9.1  &  $<30$ & 21\,53\,36.9 $+$17\,41\,42 & 21\,53\,36.81 $+$17\,41\,43.5 & 21\,53\,36.74 $+$17\,41\,44.2 & A\,2390 cD$^{f}$\cr
SMM\,J14010+0252   & ~5.4  &  ~20   & 14\,01\,02.3 $+$02\,52\,40 & 14\,01\,02.09 $+$02\,52\,42.6 & 14\,01\,02.11 $+$02\,52\,43.1 & A\,1835 cD$^{f}$\cr
\noalign{\smallskip}\hline
\noalign{\smallskip}
\end{tabular}
\end{center}
\begin{center}
$a$) Ivison et al.\ (1998a) --
$b$) Frayer et al.\ (2000) --
$c$) Smail et al.\ (1999a) --
$d$) Ivison et al.\ (2000a) --
$e$) Soucail et al.\ (1999) -- \\
$f$) Edge et al.\ (1999) --
$g$) Falls outside the 450-$\mu$m map.
\end{center}
\end{table*}
}

%
%
\begin{figure}
\centerline{\psfig{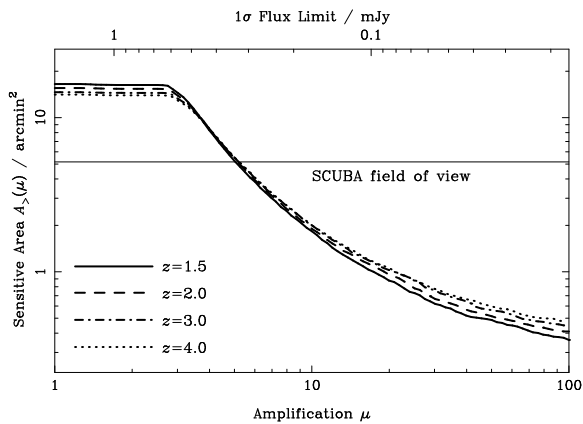} }
\noindent{\small\addtolength{\baselineskip}{-3pt}}
\caption{The total area surveyed in the source plane across all
seven clusters in our sample as a function of lens amplification and
amplification-corrected 850-$\mu$m flux density.  The latter assumes
a typical value of the effective noise in our maps of 1.7\,mJy rms. The cluster
lenses provide an increasing sensitivity across a decreasing area of the
background plane. The effect is mildly dependent on the source redshift.
Details of the magnified area for each cluster can be found in
Fig.~1 of Blain et al.\ (1999a).
}
\end{figure}

\section{Sub-mm source counts and the extragalactic background}

Our complete 850-$\mu$m sample comprises a total of 17 galaxies,
two of which are identified with the central cluster galaxies (cDs)
in the clusters A\,1835 and A\,2390. Edge et al.\ (1999) provide more
discussion of the cDs; we list their properties in Tables~2, 4 and 5 for
completeness, but exclude them from our subsequent analysis. Initially, a
further two submm sources were associated with galaxies in the foreground
of the lensing clusters (Smail et al.\ 1998); however, additional data
has provided more plausible counterparts in the background (\S5). Thus
all 15 of the non-cD detections appear to be background galaxies.

The analysis of these 15 sources makes use of well-constrained lens models
for all the clusters to accurately correct the observed source fluxes 
for lens amplification.  We use the same approach as that employed by
Blain et al.\ (1999a), but include SMM\,J04431+0210 and SMM\,J09429+4658
in the analysis, based upon the subsequent identifications of their
counterparts as background galaxies. We briefly describe the method
below. The interested reader should refer to Blain et al.\ (1999a) for
more details and references for the lens models employed for specific
clusters.

In each cluster we use the appropriate {\sc lenstool} models (Kneib et
al.\ 1996), to map the detected sources from their observed positions
back onto the source plane assuming source redshifts of $z_s = 1.5$, 2,
3 and 4 (this correction varies only slightly for sources at $z_s \gs
1$ and so we have adopted the $z_s=3$ case, in line with the results
discussed in \S4.6). The observed flux densities are also corrected for
lens amplification, leading to true flux densities in the source plane
that are less than the observed values.  Note that the amplifications
quoted in Table~4 are calculated using the redshift constraints listed
in that table, while the calculation here assumes a median redshift for
the sample as a whole.  None of the SCUBA sources appear to be multiple
images (van der Werf et al.\ 2001b) of the same galaxy (although see
SMM\,J00266+1710 in \S5) and so we count them individually. The number
counts, $N_{\rm raw}(>S, z)$, are calculated by simply summing the number
of sources brighter than a flux density $S$ {\it in the source plane,
after correcting for lensing}. A simple approximation to the Poisson
uncertainty $1.1 \sqrt {N_{\rm raw}-1}$ is attached to this value for
$N_{\rm raw} \geq 4$; for $N_{\rm raw} < 3$, results from Gerhels (1986)
are used. Poisson noise due to the modest sample size dominates both
calibration uncertainties and uncertainties in the lensing correction.

Next we estimate the area surveyed by our observations.  The area of
background sky within which a galaxy would be detectable above a given
flux limit in each cluster was determined from a map of the amplification
in the source plane, derived using the {\sc lenstool} models. Based on
this, the area of the source plane behind each cluster that lies within
the SCUBA field of view, and is amplified by a factor greater than $\mu$,
$A_>$ is shown in Fig.\,3. Due to the amplification, $A_>$ is smaller
than the SCUBA field of view.  The uncertainty in the lensing correction
of the effective area of the survey is much smaller than the Poisson
uncertainty on the counts. We have verified that this is case by assuming
different source redshifts between $z=1.5$ and 4, as shown in Fig.\,3.

To calculate the effective area of our survey to a given sensitivity limit
we use the fact that a galaxy with a flux density $S$ in the source plane
will appear in the image plane of a particular cluster above a detection
threshold $S_{\rm min}$ if it is amplified by a factor greater than $\mu =
S_{\rm min}/S$. The area in the source plane within which such a galaxy
would be detected in that cluster is then $A_>(S_{\rm min}/S, z)$. The
flux density threshold $S_{\rm min}$ and the form of $A_>$ are different
for each cluster.  By dividing the number of detected galaxies in the
catalogue $N_{\rm raw}$ by the sum of the areas $A_>(S_{\rm min}/S, z)$
for all seven clusters, the count $N(>S, z) \simeq N_{\rm raw}(>S, z)
/ \sum A_>(S_{\rm min}/S, z)$ is derived.

The final 850-$\mu$m counts are shown in Table~3 and Fig.~4. These are
within the 1-$\sigma$ errors of our previous analysis of the 850-$\mu$m
counts (Blain et al.\ 1999a).  The results are thus insensitive to
the precise identification of individual submm sources in our survey.

The effects of uncertainties in the redshift distribution of the sources
for the lensing reconstruction, and hence for the final counts, were
quantified by adopting different assumed source redshifts, as illustrated
in Fig.~2 of Blain et al.\ (1999a). The uncertainty is less than about
10\,per cent, and is dominated by the Poisson noise. An appropriate
systematic uncertainty term is included in the counts listed in Table\,3
and shown in Fig.~4.  The robustness of our amplification correction was
verified by extensive Monte-Carlo simulations (see Blain et al.\ 1999a).
We include the uncertainties associated with our lensing analysis in
the final error quoted on the counts listed in Table~3.  The total
uncertainty in the lensing correction is at most comparable to the
typical 10-per-cent error in the absolute flux calibration.

We have applied the same methods to the catalogue constructed from the
450-$\mu$m maps of the seven lensing clusters observed in the survey
(Blain et al.\ 2000).  The 450-$\mu$m counts that result from the analysis
are listed in Table~3 and Fig.~4.

The 15 sources detected at 850\,$\mu$m in this survey suffer a range of lens
amplifications, 1.5--$\gs 4$ (Table~4).  For the median source
amplification of $\sim 2.5$  (Fig.~3) our survey covers an area in the
source plane equivalent to $\sim 15$\,arcmin$^2$ to a 3-$\sigma$ flux
density limit of about 2\,mJy at 850\,$\mu$m.  At higher
amplifications, the survey covers a smaller region at a correspondingly
higher sensitivity, in particular regions of very high amplification
lie within our survey areas along critical lines.  As a result, we are
able to constrain the general form of the integral counts at even
fainter limits, down to $\ls 0.5$\,mJy, because we know the area in the
source plane over which such faint sources could be detected, and the
number of sources brighter than this limit that were detected.

The absolute limit to the depth of the survey can be determined from
Fig.~3, which shows that a significant area behind the cluster lenses
is surveyed down to the 0.25\,mJy 3-$\sigma$ flux density level --
approximately 0.7\,arcmin$^2$, or about 10 observing beams on the sky.
Note that the lens amplification in the cluster fields also results
in an effective beam size that is about half the size of the apparent
beam, enabling a fainter confusion limit to be reached as compared with
blank-field observations (see \S3.1 and Fig.~5).  However, fainter than
about 0.25\,mJy the effects of source confusion, and the uncertainty in
the area of the high-magnification region shown in Fig.~3 are likely to
become significant, and so we plot the derived counts from the direct
inversion technique as an upper limit at 0.25\,mJy.  This limit is
consistent with the lack of any clear candidates for multiply-imaged
submm sources amongst the sources with typical fluxes of 4--5\,mJy in the
catalogue. The magnification of such multiply-imaged systems is expected
to be in excess of 10, and so it is reasonable to quote only an upper
limit at a flux density of 0.25\,mJy, which is a factor of 20 deeper
than the faintest source. We note, however, that this constraint may be
strengthened by future submm surveys of clusters lenses.  In particular,
van der Werf et al.\ (2001b) have recently detected a multiply-imaged,
and thus highly amplified, submm source in A\,2218.

While both the results of the direct inversion technique and the Monte
Carlo method are of course uncertain, due to the unknown redshifts of a
fraction of the sources -- they give consistent results at flux
densities fainter than 1\,mJy (Fig.~4).  Both techniques suggest that
at least one, and probably several, of the four sources with lower
limits on their amplifications actually arises from the 0.5--1\,mJy
population.  Correlation of fainter features in the submm maps with
multi-waveband data in the radio, X-ray and mid-IR, can be used to
identify examples of the sub-mJy submm population (Ivison et
al.\ 2000a; Fabian et al.\ 2000).

Our Monte Carlo method (Blain et al.\ 1999a) yields a count of the form
$N(>S) = K (S/S_0)^\alpha$ as a function of flux density $S$.  At
450\,$\mu$m, $K = 530 \pm 300$\,deg$^{-2}$ and $\alpha=-1.8\pm0.5$ for
$S_0=20$\,mJy. At 850\,$\mu$m, $K = 3900 \pm 1300$ and
$\alpha=-1.4\pm1.0$ for $S_0=2$\,mJy. The envelopes of these results
are shown by dashed lines in Fig.~4, which show that an approximate
power-law count provides a reasonable description of the cumulative
count in the flux density ranges plotted. The equations describing
these lines at 450 and 850\,$\mu$m both have $\alpha=-1.6$. For
$S_0=25$\,mJy at 450\,$\mu$m, $K=480$\,deg$^2$; for $S_0=3$\,mJy at
850\,$\mu$m, $K=2000$\,deg$^2$.

%
%
{\small
\setcounter{table}{2}
\begin{table}
\caption{Submm counts at 450 and 850\,$\mu$m. The third column lists
the results from the current analysis. The fourth column lists the
earlier results from Blain et al.\ (1999a).}
\begin{center}
\begin{tabular}{cccc}
\hline
Band & Flux & Cumulative Count &
Blain et al.\ (1999a) \\
($\mu$m) & (mJy) & (10$^3$\,deg$^{-2}$) & (10$^3$\,deg$^{-2}$) \\
\noalign{\smallskip}\hline
\noalign{\smallskip}
450\,$\mu$m & 10.0 & $2.1 \pm 1.2$   & ...\\
            & 25.0 & $0.5 \pm 0.5$   & ...\\
\noalign{\smallskip}
850\,$\mu$m & 0.25 & $51 \pm 21$     & ... \\
            & 0.5  & $27 \pm 10$      & $22 \pm 9$ \\
            & 1.0  & $9.5 \pm 3.4$   & $7.9 \pm 3.0$ \\
            & 2.0  & $2.9 \pm 1.1$   & $2.6 \pm 1.0$ \\
            & 4.0  & $1.7 \pm 0.8$   & $1.5 \pm 0.7$ \\
            & 8.0  & $0.90 \pm 0.58$ & $0.8 \pm 0.6$ \\
            & 16.0 & $<0.42$ & ... \\
\noalign{\smallskip}\hline
\end{tabular}
\end{center}
\end{table}
}

%
%
\setcounter{figure}{3}
\begin{figure*}
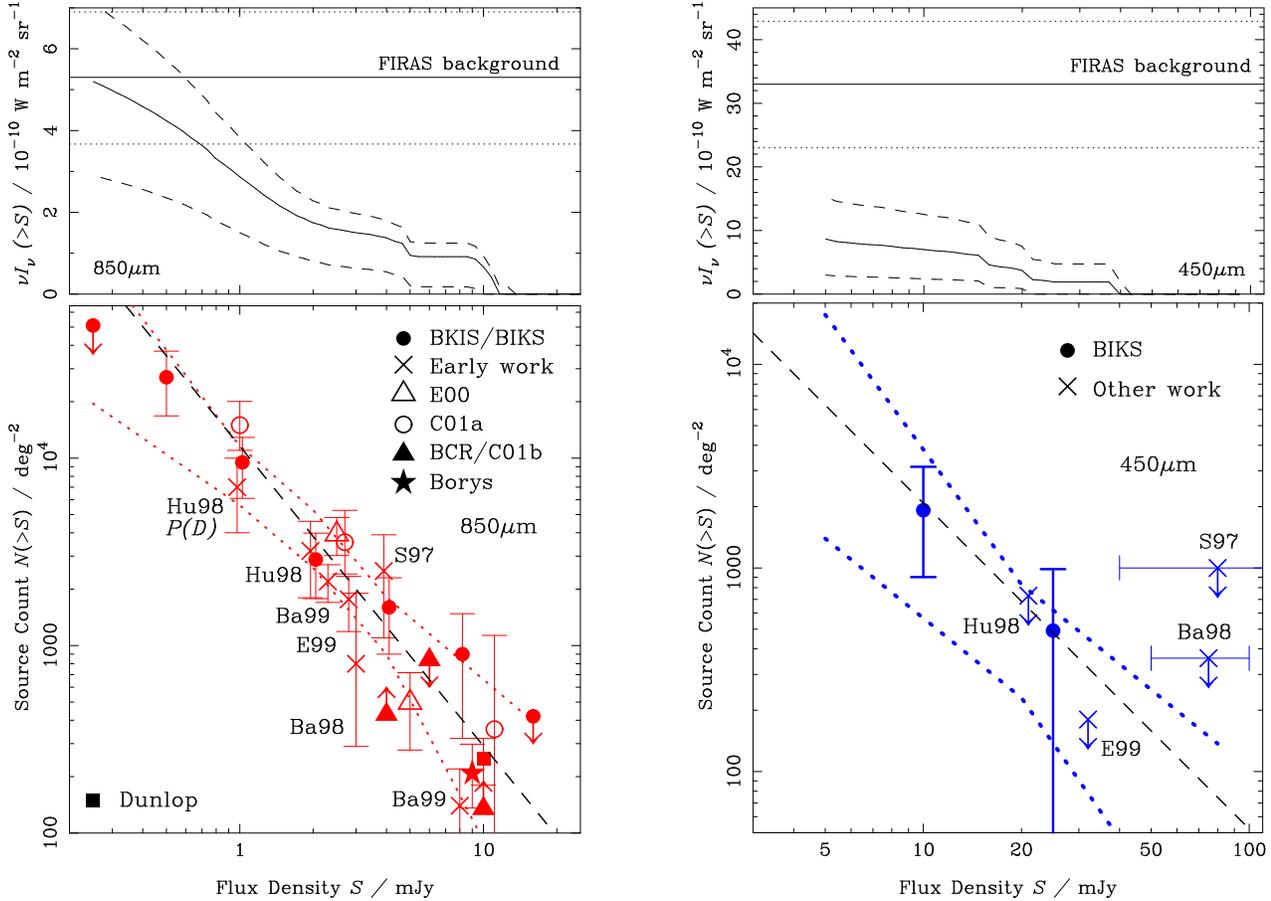

\centerline{\psfig{file=f4a.ps,angle=270,width=3.0in}
\hspace*{0.5in} \psfig{file=f4b.ps,angle=270,width=3.0in}}
\centerline{\psfig{file=f4c.ps,angle=270,width=3.0in}
\hspace*{0.5in} \psfig{file=f4d.ps,angle=270,width=3.0in}}
\caption{{\bf Left:} the 850-$\mu$m counts of galaxies, including the
updated SCUBA Lens Survey counts (corrected for lens amplification). To
avoid complicating the figure, the direct counts obtained by Blain et
al.\ (1999a; see Table~3) are not shown.  The associated Monte-Carlo
results are shown by the dotted lines.  Ba98/Ba99 -- Barger et al.\
(1998, 1999a); E99 -- Eales et al.\ (1999); Hu98 -- Hughes et al.\
(1998); S97 -- Smail et al.\ (1997). The latest counts from Eales et
al.\ (E00, 2000), Borys et al.\ (2001), Chapman et al.\ (C01a, 2001a)
and Dunlop (2001) are also shown. Limits from submm-wave
observations of infrared-faint radio-selected objects (Barger, Cowie \&
Richards -- BCR -- 2000; Chapman et al. -- Co1b -- 2001b) are also
included.  To avoid overcrowding only the faintest and brightest
results from non-cluster surveys are plotted on the figure. The top
panel compares the cumulative flux in the resolved population from a
direct integration of our counts to that estimated by {\it COBE}.  {\bf
Right:} the equivalent 450-$\mu$m counts of galaxies. The direct and
Monte Carlo counts derived here are shown by the solid points and
dotted lines respectively. The dashed line gives the parametric fit to
the counts quoted in the text. Again the top panel shows the flux of
the resolved population compared to the estimate of the total
background given by {\it COBE}.
}
\end{figure*}

We estimate that the flux density in the resolved submm population down
to 1\,mJy at 850\,$\mu$m amounts to a background radiation intensity
$I_\nu = (3\pm 1)\times 10^{-10}$\,W\,m$^{-2}$\,sr$^{-1}$, while at
450\,$\mu$m, $I_\nu = (7\pm 4)\times 10^{-10}$\,W\,m$^{-2}$\,sr$^{-1}$
down to 10\,mJy. Note that this procedure simply involves adding the
detected flux densities of the non-cluster galaxies in our survey.
Gravitational lensing does not modify surface brightness, and so there
is no need for any amplification correction. The uncertainties in this
procedure are entirely due to the limited size of the sample, the
calibration uncertainties in the SCUBA images and cosmic variance.
Compared to the total intensity of extragalactic background radiation
at these two wavelengths, measured using {\it COBE}-FIRAS (Fixsen et
al.\ 1998), we resolve about 60 and 15 per cent of the 850- and
450-$\mu$m backgrounds down to flux densities of 1\,mJy and 10\,mJy
respectively (see Fig.~4).  Unless the counts flatten rapidly below
1\,mJy, the bulk of the 850-$\mu$m background will be resolved
at flux limits of 0.1--0.5\,mJy (Hughes et al.\ 1998).

A typical galaxy in our sample has an intrinsic 850-$\mu$m flux density
of 3--4\,mJy after correcting for lensing (see Table~4). From Fig.~1,
it is clear that if these galaxies lie at $z \gs 1$, and provided
that their temperatures are not substantially less than 40\,{\sc
k}, then all are ULIRGs, with far-IR luminosities $L_{\rm FIR} \gs
10^{12}$--$10^{13}$\,L$_\odot$. There is no evidence that either of these
conditions is breached for any of our submm-selected galaxies.  Even in
low-redshift low-luminosity {\it IRAS} galaxies for which submm data is
available (Dunne et al.\ 2000; Lisenfeld, Isaak \& Hills 2000), the mean
temperature is 36\,K (but see Dunne \& Eales 2001), and there is a general
trend for more luminous objects to have higher dust temperatures. Thus
we can conclude that at least half of the submm background is produced
by ULIRGs ($L_{\rm FIR} \geq 10^{12}$\,L$_\odot$, equivalent to flux
densities $\ge 1$\,mJy), with the bulk of the remainder coming from LIRGs
(Sanders \& Mirabel 1996; $L_{\rm FIR}\geq10^{11}$\,L$_\odot$).  Hence,
most of the submm background is produced by galaxies with a relatively
narrow range of luminosities, in contrast with the very wide luminosity
range contributing to the optical and near-IR backgrounds.

\subsection{Confusion in SCUBA surveys}

Unresolved and undetected faint sources in the large SCUBA beam
generate an additional source of noise in survey maps. Estimates of
the magnitude of this effect were presented by Blain et al.\
(1998), based on the results obtained early in the SCUBA Lens Survey.
The noise properties of the SCUBA map of the HDF (Hughes et al.\ 1998)
clearly shows a non-Gaussian tail characteristic of this effect,
illustrated in their Fig.~5.

Here we simulate confusion for SCUBA by drawing a large number of
simulated background galaxy distributions from the counts
shown in Fig.~4. We then simulate the observation of these fields
using the standard $-0.5$/+1/$-0.5$ SCUBA chopping scheme and resolution.
The distribution of measured values in the absence of additional
instrumental and sky noise is shown in Fig.~5a, while the effects
of adding additional Gaussian noise with a variance of 1.7\,mJy,
characteristic of the effective noise level in the Lens Survey, is shown in
Fig.~5b.  The smooth Gaussian profile superimposed on Fig.~5a has the
same width as the estimate of confusion noise (0.44\,mJy) presented by
Blain et al.\ (1998). In Fig.\,5b the smooth profile represents a
Gaussian of width 1.7\,mJy, describing the non-confusion noise. Note that 
the distribution of flux densities shown should apply to every point 
in the sky, both those containing detected sources and those without. 

The non-Gaussian properties of the confusion signal are clearly
visible in Fig.~5a.  The core of the distribution is well described by
a Gaussian, but there is a long high-flux tail that is not acccounted
for by this distribution. The distribution is also noticeably
asymmetric, although the asymmetry is reduced by the three-position
chopping scheme.  A log-normal distribution can provide a good
description of the high-flux tail; however, it is clear from the
distribution of simulated measurements that confusion is unlikely to
be severe at flux densities greater than 2\,mJy. In Fig.~5b, the
additional noise sources swamps the confusion signal, and the noise
properties are almost Gaussian.

%
%
\setcounter{figure}{4}
\begin{figure}
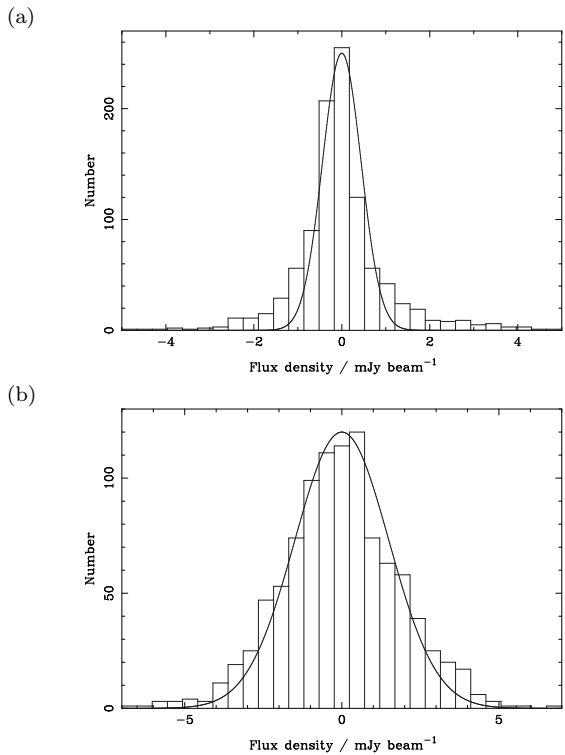

(a)
\centerline{\psfig{file=f5a.ps,width=2.5in,angle=-90}}
(b)
\centerline{\psfig{file=f5b.ps,width=2.5in,angle=-90}}
\caption{Simulated levels of source confusion in SCUBA mapping
observations.  In (a), the asymmetric confusion noise distribution in
chopped observations in the absence of instrumental or sky noise are
shown by the histogram. The smooth curve has a width corresponding to
the estimate of confusion noise in Blain et al.\ (1998). There is an
additional non-Gaussian tail to the distribution. In (b), the results
of similar simulations are shown, with an additional 1.7-mJy Gaussian
noise source added, typical of the depth of the SCUBA Lens Survey
observations. The effect of confusion is very small, as the solid
curve representing the additional sky and instrument noise is very
similar to the histogram.  }
\end{figure}

Recently, Hogg (2001) simulated the effects of confusion for reliable
determination of source positions and fluxes in 850-$\mu$m SCUBA
surveys, arguing that for reliable detections, a limit of about 2\,mJy
is required for a count with $N(>S) \propto S^{-1.5}$, in agreement with
the result above.  In contrast, Eales et al.\ (2000) have argued that
confusion biases the positions and especially the fluxes of sources at
brighter levels, $\gs 3$--4\,mJy.  The simulations of Eales et al.\
(2000) appear to assume a very steep integral count slope of $N(>S)
\propto S^{-2.6}$, based on their Fig.\,3, than found here. This count
is significantly steeper than the envelope of all the current counts
shown in Fig.\,3. Because the effects of confusion are more severe for
steeper counts (see Figs\,2,3 \& 4 in Hogg 2001), this assumption of
Eales et al.\,'s simulations is likely to account for the difference
with our results. The estimates of the positional uncertainty of sources
due to confusion, from both the Eales et al.\ and Hogg simulations,
confirm that sources can easily move by at least half the beamwidth due
to shifts in image centroids imposed by underlying confusing sources.

Based on the counts shown in Fig.\, 3, at the 3--4\,mJy depths involved
in the SCUBA Lens Survey, and the CFRS survey, confusion is unlikely to
be significant compared with instrument and sky noise, unless the data
are smoothed. The typical effect of confusion at these flux densities
is to disperse measured source flux densities by of order 0.5\,mJy,
slightly increasing the number of formal detections at the faintest
levels.  Neither of these effects would significantly modify our
results, and would certainly not lead to a uniform boosting of the flux
density of all the detections by 40\,per cent as claimed by Eales et
al.\ (2000). No systematic boost of the flux density of sources at all
observed fluxes was found here or in the results shown in Fig.\,3 of
Hogg (2001). However, for very steep counts, which may well occur at
flux densities in excess of 20\,mJy, then significant flux boosting of
this kind can certainly occur. The effects of confusion in the existing
SCUBA surveys should be easy to quantify when accurate flux densities
and positions for SCUBA sources are determined using submillimeter
interferometers such as the forthcoming Submillimeter Array (SMA; Ho et
al.\ 2000).

\section{Identifying counterparts to submm sources}

Having resolved the majority of the 850-$\mu$m background we can now move
on to attempt to study the nature of the populations contributing to it
and so determine at what epoch the background was emitted.  Here again our
survey has the advantage of lens amplification, this time in the radio and
optical/near-IR where the identification and spectroscopic follow-up are
undertaken.  Typically the counterparts of our submm sources will appear
$\sim 1$\,mag.\ brighter than the equivalent galaxy in a blank field,
providing a gain equivalent to moving from a 10-m to a 16-m telescope.

However, the low spatial resolution of SCUBA on the JCMT, 15$''$
at 850\,$\mu$m, combined with the faintness of the counterparts to
the submm sources in most other wavebands, can make identification
of counterparts on the basis of positional coincidence difficult.
The absolute astrometry of the submm maps is not an issue with these
comparisons -- it has been confirmed for several of the brighter sources
using mm-wave interferometry (Frayer et al.\ 1998, 1999, 2000; Ivison
et al.\ 2001b; Kneib et al.\ 2002; \S4.3; see also Downes et al.\
1999; Gear et al.\ 2000; Bertoldi et al.\ 2000; Lutz et al.\ 2001).
These observations confirm that the typical random astrometric errors
for bright submm sources is $\ls 2$--3$''$, this is expected to degrade
to $\sim 4''$ for the faintest sources in our sample (see Hogg 2001).

In the remainder of this section we chart the chronological development of
our source identification process to illustrate how it has evolved
over the lifetime of the project.

\subsection{Optical counterparts}

Although the submm galaxies are expected to be very dusty, it is not
foregone conclusion that they will be either very red or optically faint
-- the restframe UV colours of local ULIRGs span a wide range and include
examples of relatively blue galaxies (Trentham et al.\ 1999; Fig~8).
We therefore began our search for counterparts of the submm sources in
our survey in the optical waveband.

One of the key features used in selecting the SCUBA survey clusters for
this study was that they were well studied, and in particular that
deep, high-quality archival optical imaging existed for each of the
fields, including {\it Hubble Space Telescope} ({\it HST}\,) {\it
WFPC2} imaging. 

Details of the optical observations, their reduction and analysis can be
found in Smail et al.\ (1998).  More detailed discussion of the optical
observations of specific fields is given by Ivison et al.\ (1998a) and
Soucail et al.\ (1999) for A\,370 by Ivison et al.\ (2000a, 2001b) for A\,1835
and by Smail et al.\ (1999a) for the MS\,0440$+$02 and Cl\,0939$+$47 fields.

All the frames are calibrated to provide standard Cousins $I$-band
photometry and are linked astrometrically to the APM coordinate system
with a typical accuracy of $\ls 0.5''$.  Colours are measured using
$2''$-diameter apertures from the seeing matched $I$- and $K$-band
frames (Table~4).  When total $I$-band magnitudes are quoted these refer to
$4''$-diameter aperture photometry corrected for light falling outside the
aperture assuming a point source profile.

We show in Fig.~6 the submm maps of each of the 15 cluster background
sources overlayed on the deep $I$-band exposures of these frames.  For
the sources in A\,2390, Cl\,2244$-$02 and most of those in Cl\,0024+16,
we use {\it HST} F814W ($I$) images degraded to the same seeing and
pixel scale as ground-based near-IR images (see \S4.2).  For the
remaining fields we have used deep, ground-based $I$-band frames: for
Cl\,0939+47 and MS\,0440$+$02 we use the Keck $I$-band imaging from
Smail et al.\ (1999a) and for A\,1835 and A\,370 the Hale 5-m and CFHT
imaging discussed in Smail et al.\ (1998).

However, in the course of this analysis it has become clear that
identifying submm counterparts using optical data {\it alone} is a
problematic process (Smail et al.\ 1998; Lilly et al.\ 1999).  With the
exception of a handful of three unusual and optically-bright counterparts:
SMM\,J02399$-$0136 (Ivison et al.\ 1998); SMM\,J02399$-$0134 (Soucail
et al.\ 1999) and SMM\,J14011+0252 (Ivison et al.\ 2000a), the majority
of the submm sources cannot be reliable identified on the basis of just
optical imaging, irrespective of its depth (\S4.6; Smail et al.\ 2000).

\subsection{Near-infrared counterparts}

To more reliably identify counterparts to the 15 submm sources we next
obtained near-IR imaging of our fields. The goal is to combine this with
the deep optical data, to attempt to locate any counterparts within the
submm error box on the basis of their unusual optical--near-IR colours,
e.g.\ $(I-K)\gs 5$. For this purpose the depth required in the $K$-band 
was set by the depth of the available $I$-band images, $I\sim 25$--26,
with the deepest being the multi-orbit {\it HST} exposures, leading to a
limit of $K\sim 20$--21 for our observations.  In most cases, integrating
fainter in $K$ might produce additional candidate counterparts but we
would be unable to identify these as unusual on the basis of their very
red colours from our existing optical images.  In the few cases where
more accurate positional information about a probable submm source is
available, from either mm-wave continuum or radio interferometry maps,
deeper $K$-band observations have been obtained (e.g.\ Ivison et al.\
2000a; Frayer et al.\ 2000).

The near-IR observations of our fields were undertaken in
typically good conditions during several observing runs in late 1998
and early 1999 using the IRCAM3 and UFTI cameras on the 3.8-m
UKIRT\footnote{UKIRT is operated by the Joint Astronomy Centre on
behalf of the Particle Physics and Astronomy Research Council of the
United Kingdom.}.  Observations consist of deep $K$-band exposures
(with $J$- or $H$-band observations of the brighter sources detected
in $K$).

Data were obtained with IRCAM3 on the nights of 1998 July 11--16 and
18, September 10 and 19, October 9 and 1999 February 8--12.  The new
UFTI camera, which provides finer sampling of the average seeing on
UKIRT, was used on 1998 October 18--19 and November 19.  The seeing
during these nights was typically 0.4--0.8$''$.  Data taken in
non-photometric conditions (1998 July 16, September 19 and October
18--19) were calibrated using independent exposures taken on
photometric nights.  Total exposure times for the individual fields
are listed in Table~4.

Observations of the MS\,0440$+$02 and Cl\,0939$+$47 fields are
discussed in Smail et al.\ (1999a), while Ivison et al.\ (1998a, 2000a)
discuss the A\,370 and A\,1835 fields, respectively. Frayer et
al.\ (2000) present very deep near-IR observations of
SMM\,J00266+1708.  Here we describe the results for the complete
sample.  We note that a programme of deeper near-IR imaging and
spectroscopy, targeting the brightest submm sources, is currently
underway on Keck to attempt to identify faint near-IR counterparts in
some of the blank or ambigious fields (see Frayer et al.\ 2000 for the
first results).

The UKIRT observations used repeated sequences of nine exposures
dithered on a $3\times 3$ grid with a $10''$ separation to allow the
science images to be flatfielded using a running median sky frame
constructed from the adjacent science frames.  The frames were
linearised, dark subtracted, flatfielded and combined in a standard
manner.  The final stacked frames were photometrically calibrated using
UKIRT Faint Standards (Casali \& Hawarden 1992).  The astrometry on
these frames comes from comparison with the $I$-band exposures of these
fields.

We show in Fig.~6 $K$-band images of the fields of the sources and
compare these to the $I$-band images to illustrate the colours of possible
counterparts.  In combination with the deep optical data, it is simple
to pick out several unusually red galaxies, which are bright in the
$K$-band and faint or undetected in the $I$-band.  Of these, amongst the
most striking are the extremely red objects (EROs, e.g.\ Smith et al.\
2001b) in the fields of SMM\,J09429+4658 and SMM\,J04431+0210 (\S5).
The low surface density of such very red galaxies in blank fields,
combined with the possible connection between their extreme colours and
the presence of dust, both argue for these galaxies being the correct
counterparts to the submm sources (see Smail et al.\ 1999a).

~From the near-IR imaging of our complete submm sample, which reaches a
median 2$\sigma$ depth of $K\sim 21$ ($K\sim 22$ in the source plane),
we identify counterparts to about half of the submm sources in the
sample, of these around half again are also visible in the optical.
Table~4 lists photometry for the various counterparts and we discuss
these further in \S5.  In Table~4, a non-detection corresponds to the
$2\sigma$ limit on the total magnitude of a point source.

\subsection{Radio counterparts}

Deep radio maps provide the most powerful complement to our submm
observations.  This is particularly true of observations at 1.4\,GHz
from the NRAO Very Large Array\footnote{The National Radio Astronomy
Observatory is a facility of the National Science Foundation, operated
under cooperative agreement by Associated Universities, Inc.} (VLA).
The resulting maps are sensitive to star-forming galaxies out to high
redshifts (e.g.\ Richards 1999; Smail et al.\ 1999b), have a spatial
resolution that is well matched to the expected size of distant galaxies,
and high astrometric precision; they thus provide the best opportunity to
unambigiously identify the counterparts of submm sources (Ivison et al.\
1998a, 2000a, 2001b).

The deep 1.4-GHz maps used by Smail et al.\ (2000) to identify radio
counterparts for the submm sources in our sample were all obtained with
the VLA in A and/or B configuration, giving effective resolutions of
1--5$''$. In A configuration, most of the detected galaxies are
resolved, some more than others: for SMM\,J02399$-$0136 at $z=2.80$ the
two interacting/merging components, L1 and L2, are resolved from each
other, and resolved individually (Ivison et al.\ 1999; Owen et al., in
prep). While for SMM\,J14011+0252 the combined A+B configuration map
constructed by Ivison et al.\ (2001b) resolves the starburst region and
shows that it is distinct from the optically-luminous components of
this system.  In contrast the B configuration maps may provide a more
reliable measure of the total radio flux in these extended systems, and
hence a better comparison to large-aperture submm fluxes.  We list the
total exposure times and 1-$\sigma$ sensitivity limits of the maps in
Table~1.  The typical 3-$\sigma$ sensitivity of these deep maps on the
background source plane is about 18\,$\mu$Jy, and so the sensitivity is
sufficient to detect a ULIRG at a redshift $z\simeq 3$ (Downes et
al.\ 1999).

A summary of these observations is given by Smail et al.\ (2000). More
details of the reduction and analysis of individual maps can be found
in the following references: for Cl\,0024$+$16 in Morrison (1999);
for A\,370 in Ivison et al.\ (1998a); for MS\,0440+02 in Smail et al.\
(1999a); for Cl\,0939+47 in Smail et al.\ (1999a, 1999b); for A\,1835 in
Ivison et al.\ (2000a, 2001b); and for A\,2390 in Edge et al.\ (1999).
Note that the bright radio source associated with the central cluster
galaxy in A\,2390 reduces the sensitivity of the map of this field.

Smail et al.\ (2000) searched for radio counterparts within the nominal
error boxes of the submm sources.  In ambigious cases where the nominal
error box is blank in the radio, but a radio source is visible just
outside, they adopted the conservative approach of using the flux of that
source as a limit on the radio counterpart.  When using the radio and
submm fluxes to place limits on the possible redshifts of the sources,
this approach leads to a firm lower redshift estimate for the submm source
(see \S4.6).

Using the radio maps, Smail et al.\ (2000) found counterparts to roughly
half the sample (Table~4). Several of these correspond to previously
identified counterparts from the optical or near-IR identification
campaigns. However, a number were new identifications and these prompted
further, deeper near-IR observations using the improved positions from
the radio maps to attempt to detect counterparts (e.g.\ SMM\,J14009+0252,
Ivison et al.\ 2000a).

We list the apparent radio fluxes or 3$\sigma$ limits in Table~5.  For
the 10 submm sources with radio counterparts, we measure a median
positional offset of only 2.0$''$ between the submm and radio
positions.  This supports the estimates of the positional accuracy of
the submm sources, $<2$--4$''$, quoted in \S2 and used in the searches
for optical and near-IR counterparts (Smail et al.\ 1998; \S4).  We
note that the good positional accuracy of our submm sources confirms
that confusion is not a significant concern for our survey
(c.f.\ Eales et al.\ 2000).

\subsection{Mid-IR properties}

Several of the cluster lens targeted by our SCUBA survey have also been
imaged in the mid-infrared (mid-IR) using {\it ISOCAM}.  Maps of A\,370
and A\,2390 at 15\,$\mu$m are discussed by Altieri et al.\ (1999) and
Metcalfe et al.\ (1999), while similar observations have been obtained
for Cl\,0024+16 and Cl\,2244$-$02 (L.\ Metcalfe, priv.\ comm.).  In total
there are five submm sources in A\,370 and A\,2390, and if we discount the
central galaxy in A\,2390 (which is detected by {\it ISO}, Edge et al.\
1999), we find that three of the remaining four submm sources are detected
in the {\it ISO} 15-$\mu$m maps: SMM\,J02399$-$0136 (Ivison et al.\
1998a), SMM\,J02399$-$0134 (Soucail et al.\ 1999) and SMM\,J21536+1742,
at apparent flux limits of $\sim 0.2$--2\,mJy. This implies that a minimum
of 20 per cent of the submm population (and possibly up to 75 per cent)
have 15-$\mu$m counterparts.  A similar rate of correspondence between
the two populations, 20 per cent, was found by Hughes et al.\ (1998)
in the HDF, with Eales et al.\ (2000) finding a slightly lower rate
(10 per cent) in their sample.

The long-wavelength photometer, {\it ISOPHT}, was also used to produce
175-$\mu$m maps of three of the fields included in our survey: A\,370,
Cl\,2244$-$02 and MS\,0440+02 for the programme RIVISON/CLUSTERS.
Despite long exposure times and large maps, not even the brightest
sources in these fields  (e.g.\ SMM\,J02399$-$0136) were detected at
3--$\sigma$ flux limits of $\sim 150$\,mJy.

%
%
\begin{figure*}
\centerline{\psfig{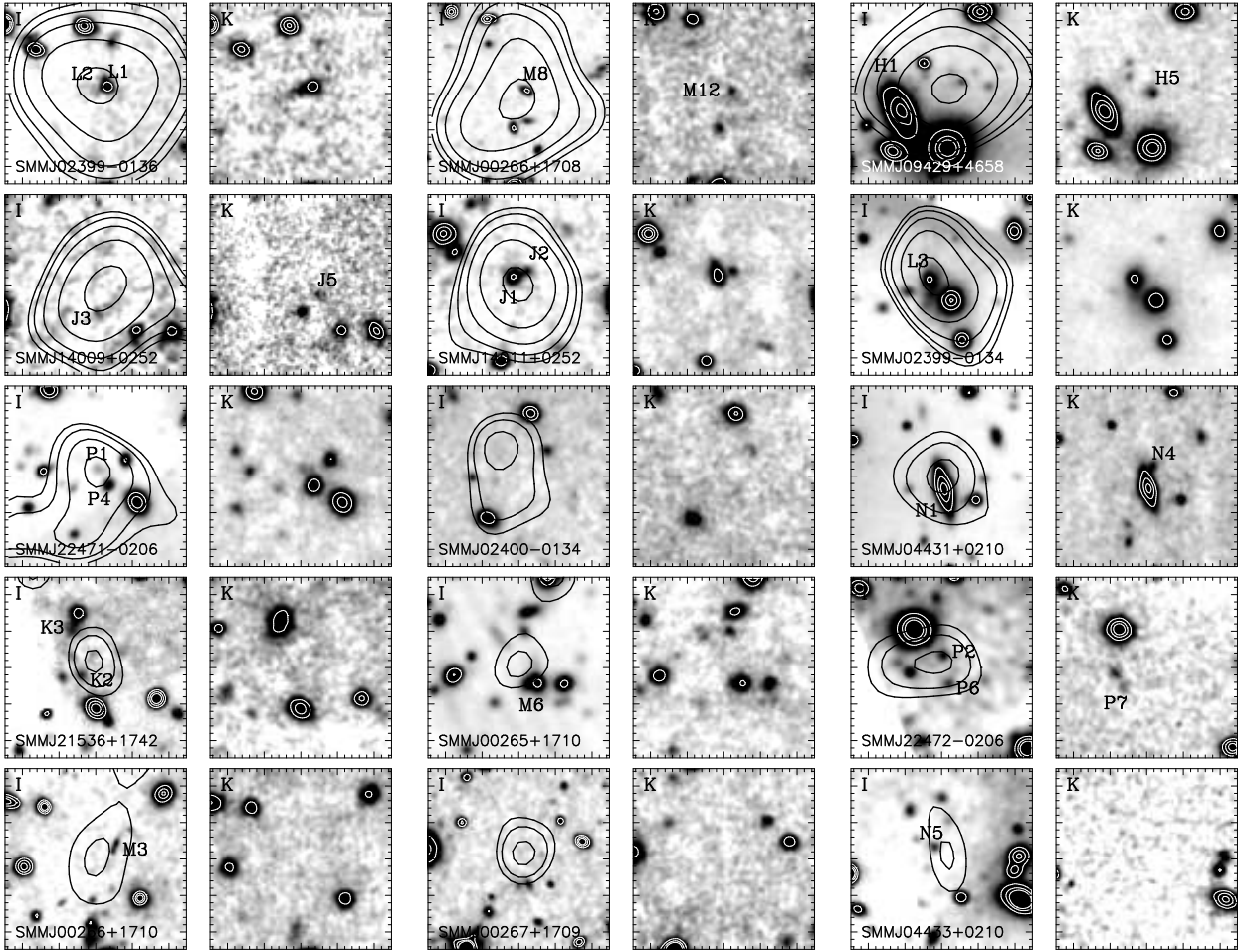}}
\caption{$I$- and $K$-band frames of the fields of the submm sources
in our sample (excluding the two central cluster galaxies), ordered in
terms of their apparent 850-$\mu$m fluxes.  The 850-$\mu$m map of each
source is overlayed as a contour plot on the $I$-band frame, after
convolution with an 8$''$-FWHM gaussian for display purposes.  Note
that the $I$- and $K$-band frames represent a range in depth and
resolution, but they have all been smoothed (with a 0.5$''$-FWHM
Gaussian) to enhance the visibility of faint features.  We identify
the various confirmed or candidate counterparts and other galaxies
discussed in the text on the individual panels.  Each panel is
25$''$-square and has north top and east to the left.}
\end{figure*}

\subsection{X-ray properties}

X-ray observations of the submm galaxies provide a useful
addition to our view of this population, in part because of
the ability of hard X-ray photons produced by AGN to escape from
obscured environments.  X-ray data can thus be used to search for
obscured AGN in the submm population which might be partly responsible
for the extreme luminosities of these sources.  For this reason there
has been considerable interest in obtaining hard X-ray observations of
the submm population (Almaini, Lawrence \& Boyle 1999; Gunn \& Shanks
2001).

The high resolution X-ray imaging capabilities of the {\it Chandra}
X-ray satellite make it well suited to tackle this problem. In
contrast, although it has a much higher sensitivity to hard X-rays, data
from {\it XMM/Newton} are less ideal for the cluster fields studied
here -- the lower spatial resolution making it difficult to detect any
emission from the submm galaxies above the strong cluster X-ray
emission. Fabian et al.\ (2000) discuss sensitive {\it Chandra}
observations of the submm sources in our A\,1835 and A\,2390 fields,
while Bautz et al.\ (2000) present somewhat deeper observations of
A\,370.  Excluding the central cluster galaxies, these observations
cover six of the submm sources in our sample; of these, only two are
detected, both of which were already known to host semi-obscured AGNs
on the basis of their optical and mid-IR emission (SMM\,J02399$-$0136
and SMM\,J02399$-$0134).

To extend this analysis to our full sample we have also made use of
deep archival {\it ROSAT} {\it HRI} images of all our fields.  Details
of the exposure times and sensitivities of the {\it ROSAT} observations
are given in Table~1. These lack both the sensitivity and coverage of
the hardest X-ray energies compared to the recent {\it Chandra}
observations; nevertheless, they do allow us to place limits on the
presence of strong X-ray emission for {\it all} of the sources in our
sample.  We find that no sources have detectable X-ray counterparts
down to 3-$\sigma$ flux limits of $\sim 3 \times
10^{-15}$\,ergs\,s$^{-1}$\,cm$^{-2}$ in the 0.1--2.0\,keV band (roughly
0.4--8\,keV in the rest frame for most of these sources) after
correcting for Galactic H\,{\sc i} absorption.  At a redshift of $z\sim
3$ this translates into an X-ray luminosity in the 2--10\,keV band of
$L_X\sim 3 \times 10^{43}$\,ergs, assuming an unobscured $\Gamma=2$
power-law SED (see the discussion in Bautz et al.\ 2000).

\subsection{Redshift constraints}

Constraining the redshifts of possible submm counterparts can also provide
a useful route to identifying the correct counterpart to the submm source.
These constraints can come  from either classical spectroscopy (Barger et
al.\ 1999a) or photometric techniques employing optical, near-IR and longer
wavelength observations (e.g.\ Hughes et al.\ 1998; Lilly et al.\ 1999;
Smail et al.\ 2000).  Moreover, determining the redshift distribution of
the submm population is also important to provide a better understanding
of the nature of these galaxies, their relationship to other classes of
high-redshift source and their importance for models of galaxy formation
and evolution. This is especially important for submm-wave surveys,
as the selection function is extremely broad, and so a priori very high
redshift galaxies could be included in the sample.

\subsubsection{Optical photometry}

Photometric redshift techniques based on optical/UV photometry have
been employed by Lilly et al.\ (1999) to attempt to determine the
redshifts of candidate optical counterparts found within the
positional error box of submm sources.  The results suggested that the
bulk of the submm population lies at $z=0.1$--3. Hughes et al.\ (1998)
used a photometric technique spanning a larger wavelength range and
deduced a range of $z=2$--4.  Such analyses are based on template
spectral energy distributions (SEDs) of local optically-selected
galaxies, as sampled in the rest-frame UV waveband if the galaxies are
at high redshifts.  The diversity of the rest-frame UV/optical
properties of even local ULIRGs (Trentham et al.\ 1999) indicates that
such analyses, which employ SED templates derived from `normal'
galaxies, are unlikely to be reliable. The SEDs of SCUBA galaxies with
known redshifts (Ivison et al.\ 1998a, 2000a; Soucail et al.\ 1999)
are also diverse and unusual, while the many very red
counterparts (Smail et al.\ 1999a; Dey et al.\ 1999; Ivison et al.\
2000b; Gear et al.\ 2000; Lutz et al.\ 2001) would not be well described by
standard photometric redshift templates.

\subsubsection{Optical spectroscopy}

If optical counterparts could be identified for most submm sources then
obviously optical spectroscopy would provide the most direct route to
obtain the redshift distribution of the population, $N(z)$. Based on
the list of plausible optical counterparts from Smail et al.\ (1998),
a spectroscopic survey was undertaken with the Keck telescope (Barger
et al.\ 1999a).

Identifications were attempted for all the galaxies bright enough for
reliable spectroscopy within the SCUBA error circles.  This resulted in
spectroscopic redshifts or limits for 24 possible counterparts to 14
SCUBA sources (SMM\,J04433+0210 and the two central cluster galaxies
were omitted from the survey). The median $I$-band magnitude of the
counterparts is $I=22.4$; the equivalent depth for identifying
candidates in a blank-field submm survey would be closer to $I\sim
23.5$ (Fig.~8), stretching the capabilities of even the largest
telescopes.  Importantly, however, Barger et al.'s survey allows us to
identify candidate counterparts containing AGN based on their spectral
properties, and thus allows an upper limit to be imposed on the
proportion of submm galaxies which host unobscured and partially
obscured AGN. One route to rule out candidate optical counterparts for
SCUBA sources is to compare their optical spectroscopic redshifts with
redshift limits obtained from their radio and submm properties.

\subsubsection{Submm colours}

In the absence of bright optical counterparts for most submm sources,
and doubts over the applicability of standard optical photometric methods
for these dusty systems, we must resort to other redshift indicators.
Proposed techniques to estimate redshifts for faint submm galaxies
come from the analysis of their long wavelength SEDs.  For example, as
discussed by Hughes et al.\ (1998), the ratio of 450- and 850-$\mu$m
fluxes, i.e.\ the spectral shape of dust emission in the rest-frame
far-IR, can be used as a crude redshift indicator.

The analysis of the information provided by the 450-$\mu$m
non-detections of the five 850-$\mu$m sources in the HDF (Hughes et
al.\ 1998) suggested that the galaxies all lie at $z>1$. A similar
constraint comes from assuming that the same population of sources are
being detected at 450 and 850\,$\mu$m, and then determining the flux
density ratio between the two wavelengths at a fixed source surface
density. That the redshift distributions of submm sources at the two
wavelengths are likely to be rather similar is supported by Eales et
al.\ (1999) and Blain et al.\ (1999b).  The 450-$\mu$m counts are
1000\,deg$^{-2}$ brighter than at a flux density limit of 20\,mJy. At
850\,$\mu$m, this surface density is reached at a flux density of
about 6.5\,mJy. Thus the typical ratio $S_{450}/S_{850} \simeq 3$,
suggesting that $<\! z\!> \sim 3$ if the dust temperature in the
population is about 40\,{\sc k}, see Fig.~1. This flux ratio is
consistent with the mean value $3.4\pm0.6$ for the five galaxies
detected at both wavelengths (see Table~2).

A more reliable estimate of the redshifts of individual sources comes from
fitting template SEDs to the entire far-IR/submm/mm SED, provided that
sufficient information is available (see for example Ivison et al.\ 2000a;
Frayer et al.\ 2000; Smail et al.\ 1999a; Fox et al.\ 2001). We list the
crude redshift ranges derived from this technique in Table~5.  However,
it is vital to recognise that these estimated redshifts depend strongly on
the SED template that is assumed. Because the dust spectrum is thermal,
there is a strong degeneracy in the results between the detection of a
cooler source at low redshift and a hotter source at a greater distance.

\subsubsection{Radio--submm spectral index}

A method has been developed recently to estimate redshifts for submm
sources exploiting measurements of the 850\,$\mu$m to 1.4\,GHz spectral
spectral index, $\alpha^{850}_{1.4}$ (Carilli \& Yun 1999, 2000; Blain
1999; Dunne, Clements \& Eales 2000; Barger, Cowie \& Richards 2000).
This technique relies upon the tight correlation between the strength
of the far-IR emission (reprocessed UV/optical radiation from massive
stars) and radio emission (synchrotron emission from electrons
accelerated in supernovae from massive stars) observed in local
star-forming galaxies (Condon 1992).

The decline in emission from dust at longer wavelengths is eventually
overtaken by the rising synchrotron emission to produce an upturn
between the submm and radio wavebands at around 3\,mm. This spectral
feature is also observed in both high-redshift AGN and star-forming
galaxies (see Fig.~10). As proposed by Carilli \& Yun (1999), the
spectral index observed across this break can be used to provide a
crude redshift estimate, with a larger spectral index indicating a
higher redshift. The spectral index has the useful property that
contamination by radio emission from an obscured radio-loud AGN will
tend to reduce the value of $\alpha^{850}_{1.4}$, and thus lead to an
underestimate of the redshift.  Carilli \& Yun (1999, 2000) were able
to show that the redshift predictions from $\alpha^{850}_{1.4}$ based
on local template spectra and model SEDs were in good agreement with
the observed redshifts for a small sample of distant submm sources.
Thus $\alpha^{850}_{1.4}$ can be used to place robust {\it lower}
limits on the redshifts of the submm population.

%
%
\begin{figure}
\centerline{\psfig{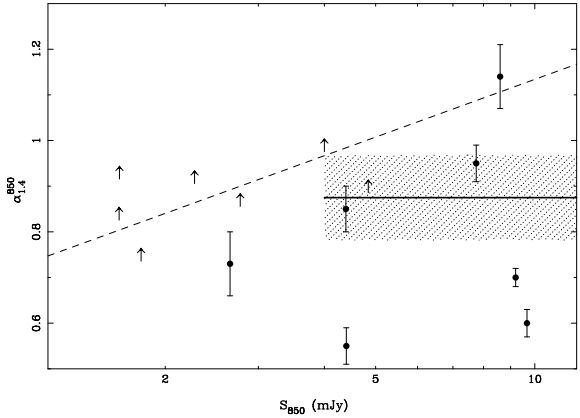}}
\noindent{\small\addtolength{\baselineskip}{-3pt}}
\caption{The distribution of radio--submm spectral indices,
$\alpha^{850}_{1.4}$, with apparent 850-$\mu$m flux for the submm
population.  The dashed line indicates the approximate source-plane
limits of our VLA 1.4-GHz maps.  The solid line and hatched region
shows the median value and 1-$\sigma$ confidence limits for the
$S_{850}\geq 4$\,mJy population.  The 850-$\mu$m fluxes and limits
have been corrected for lens amplification using the estimated
amplification factors from Table~4.  }
\end{figure}

Using the deep VLA 1.4-GHz maps of the seven clusters in our survey
(Table~1), Smail et al.\ (2000) identified radio counterparts to around
half of the submm sources, with useful limits on the remainder.  They
employed models describing the behaviour of $\alpha^{850}_{1.4}$ with
redshift from Carilli \& Yun (1999) and Blain (1999) to convert their
$\alpha^{850}_{1.4}$ measurements and limits into redshift ranges for
the galaxies.  Using the updated models from Carilli \& Yun (2000),
which include a representation of the scatter expected in
$\alpha^{850}_{1.4}$ as a function of redshift, we estimate the
redshift limits and ranges, $z_\alpha$, for the individual submm
galaxies and list these in Table~5.

Taking the various estimates of submm source redshifts listed in Table~5
we can place a crude limit on the median redshift of a complete sample
of the submm galaxy population.  We include both the reliable and
possible spectroscopic redshifts, distribute those galaxies for which
upper limits are available on $z_\alpha$ from the radio--submm spectral
indices uniformly across their allowed redshift ranges, and place the
remaining galaxies with only lower limits on $z_\alpha$ at the minimum
allowed redshift (Table~5).  In this way we estimate a conservative
lower limit of $<\! z\!>=2.3^{+0.8}_{-1.1}$ (68 per cent confidence
limits) to the median redshift of the submm population brighter than
1\,mJy, with only a modest fraction of the submm population at $z< 1$.
If we instead adopt a less conservative, more realistic, approach and
uniformly distribute those galaxies with only lower limits to $z_\alpha$
across a redshift range from $z_\alpha$ out to some maximum redshift
($z\sim 6$), we obtain a slightly higher median redshift $<\!  z\! >
\sim 2.6^{+1.9}_{-1.4}$ (see Hughes et al.\ 1998; Fox et al.\ 2001).
At these redshifts and submm fluxes, all of the galaxies in our sample
must be either ultraluminous or hyperluminous infrared galaxies, $L_{\rm
FIR} \sim 10^{12}$--$10^{13}$\,L$_\odot$.

In Figure~7 we illustrate the variation of radio--submm spectral index
with submm flux for the galaxies in our sample.  The flat
flux--redshift relation for submm galaxies shown in Fig.~1 means that
any correlation between the redshift-dependent spectral index and
850-$\mu$m flux must arise from intrinsic differences in the submm
population as a function of luminosity.  As pointed out by Smail et
al.\ (2000), there is a hint that the submm galaxies at lower fluxes
have typically fainter radio counterparts and thus higher
$\alpha^{850}_{1.4}$ values.  However, this is at the limits of the
sensitivity of even the very deep radio maps used in our analysis.  We
discuss the variation of the properties of the submm galaxies with
their submm flux, as well as the sensitivity of our conclusions on the
median redshift of the population to the detailed properties of the
submm galaxies, in \S6.2.

\section{The properties of individual submm galaxies}

In the following we give a brief summary of the properties of the
galaxies in our sample:

{\it SMM\,J02399$-$0136} -- This was the first high-redshift submm
galaxy to be detected using SCUBA and remains one of the brightest
(850-$\mu$m flux of 23\,mJy based on a weighted mean of photometry and
map measurements, Ivison et al.\ 1998a).  It was quickly identified
with a luminous interacting/merging galaxy, L1/L2 at $z=2.80$, with a
far-IR luminosity of $\sim 10^{13}L_\odot$.  Both components are
resolved in our UKIRT $K$-band image with FWHM of $\sim 0.3$--0.4$''$.
This system is a relatively bright mid-IR source seen by {\it ISOCAM}
(the flux densities have been revised to $0.13\pm 0.06$ and $0.47\pm
0.05$\,mJy at 7 and 15\,$\mu$m since those quoted by Ivison et
al.\ 1998a -- Metcalfe et al.\ 2001, in prep). The two components are
each resolved by a 1.3$''$ synthesised beam at 1.4\,GHz and appear to
have different radio spectral indices (Ivison et al.\ 1999). The
rest-frame UV spectrum exhibits high-ionisation lines with widths of
$\sim 1000$\,km\,s$^{-1}$ leading to its classification as a Seyfert-2
galaxy (see also Vernet \& Cimatti 2001).  However, the strong CO
detection of this galaxy suggests that around half of its luminosity
arises from a starburst, with the remaining half coming from the AGN
(Frayer et al.\ 1998). Moreover, these observations demonstrate the
massive and relatively unevolved nature of this galaxy, with a large
gas fraction, an SFR of $\sim 10^3 M_\odot$\,yr$^{-1}$ and an estimated
gas reservoir amounting to $M({\rm H}_2)\sim 2 \times
10^{11}$\,M$_\odot$ (this would probably be revised upwards were
CO($1\rightarrow0$) data available -- see Papadopoulos et al.\ 2001).
With this SFR, the galaxy could add another $10^{11}$\,M$_\odot$ of
stars in the next 100\,Myr to its already considerable stellar
luminosity, $L_V\sim 6 L^\ast$.  We note that the recent detection of
this galaxy in hard X-rays by {\it Chandra} confirms the presence of a
partially obscured AGN in this source and also supports the claim for a
significant starburst contribution to the total luminosity (Bautz et
al.\ 2000).

{\it SMM\,J00266+1708} -- Frayer et al.\ (2000) present the
identification of this submm source with the galaxy M11 based upon 1-mm
data from the OVRO Millimeter Array interferometer.  A faint near-IR
counterpart was identified in their deep Keck $K$-band imaging (we list
the properties of this galaxy from their observations in Table~4).  The
current limits on the $I$-band magnitude of this galaxy do not allow us
to classify it as an ERO, but clearly it is very faint and red and so
this classification is relatively academic.  Based on the
long-wavelength SED of this source, Frayer et al.\ (2000) suggest its
redshift $z\sim 3.5$ and that $L_{\rm FIR}\sim 10^{13}$\,L$_\odot$.

{\it SMM\,J09429+4658} -- This source was originally identified with
the bright spiral, H1 (Smail et al.\ 1998).  However, near-IR imaging
of this field turned up the counterpart H5, one of two bright EROs
identified in our UKIRT $K$-band survey.  This galaxy is well resolved
in the $K$-band image with a seeing-corrected FWHM of $0.5\pm 0.1''$.
This source is discussed in more length in Smail et al.\ (1999a) who
estimate a redshift of $z\sim 2.5$ for this ULIRG ($L_{\rm FIR}\sim
2\times 10^{13} L_\odot$).

{\it SMM\,J14009+0252} -- The counterpart (J5) to this submm source is
extremely faint in the near-IR (and even fainter in the optical).
However, there is relatively bright radio emission associated with
this galaxy (Table~5), unresolved in a sensitive 1.5$''$-resolution
VLA map of the field, and the accurate position allowed us to
pinpoint the position of the submm galaxy with sufficient accuracy to
warrant a deeper near-IR observation.  A near-IR counterpart was
eventually detected with $K\sim 21.0$ (Ivison et al.\ 2000a) although
it remains undetected in a very recent {\it HST} {\it WFPC2} F702W
exposure, indicating it has $(R-K)>5.6$ (Table~4) and therefore 
classes as an ERO. Balancing the high redshift suggested by the
$S_{450}/S_{850}$ flux ratio and the low redshift given by the
$\alpha^{840}_{1.4}$ estimator, Ivison et al.\ (2000a) suggest that
this source lies at $z\sim 4$ and class the galaxy as a hyperluminous
IR galaxy, $>10^{13}L_\odot$ with a probable AGN-based contribution to
its radio emission.

{\it SMM\,J14011+0252} -- Ivison et al.\ (2000a) present detailed
observations of this submm source, which is identified with a $z=2.56$
interacting/merging pair of galaxies, J1/J2, with a far-IR luminosity
of $L_{\rm FIR}\sim 6\times 10^{12}$\,L$_\odot$.  Both optical components are
resolved at $K$, J1 being larger ($1.05\pm 0.05''$) than J2 ($0.4\pm
0.1''$). The optical spectra of these galaxies show no hint of AGN
characteristics (Barger et al.\ 1999a) and the system is undetected
in recent hard X-ray observations with {\it Chandra} (Fabian et al.\
2000), supporting the contention that its luminosity is predominantly
produced by a intense starburst  (Ivison et al.\ 2000a).  The resolution
of the 1.4-GHz radio emission from SMM\,J14011+0252 in our VLA A+B-array
observations also suggests that the far-IR emission from this galaxy
is powered by a starburst (Ivison et al.\ 2001b).  Based on optical
photometry of the galaxy pair, Adelberger \& Steidel (2000) noted that
J1/J2 also classes as a Lyman-break galaxy (LBG). However, 
strong colour differences exist within this system and it is not clear that
the source of the starburst activity will have colours typical of a LBG.
Indeed, the high-resolution study of SMM\,J14011+0252 by Ivison et al.\
(2001b) suggests that the starburst region lies outside the optical
extent of the system, close to a previously unidentified extremely
red component.  The CO detection of this galaxy by Frayer et al.\
(1999), with a relatively narrow linewidth, indicates that it has a
large gas reservoir ($M({\rm H}_2)\sim 10^{11}$\,M$_\odot$) and that
the gas fraction is high (see also Ivison et al.\ 2001b).

{\it SMM\,J02399$-$0134} -- This submm source is identified with L3, a
ring-galaxy at $z=1.06$ (FWHM in $K$: $0.6\pm 0.1''$), which is also a
relatively bright VLA 1.4-GHz source and an {\it ISOCAM} 7- and
15-$\mu$m source (with a 15-$\mu$m flux density revised to $1.4\pm
0.1$\,mJy since that quoted by Soucail et al.\ 1999 --- Metcalfe et
al.\ in prep).  Optical spectroscopy (and analysis of the mid-IR
colours) shows that the galaxy is a Seyfert~1.5--2, and the presence of
an obscured AGN was recently confirmed through a detection in the hard
X-ray band by {\it Chandra} (Bautz et al.\ 2000).  The
lensing-corrected far-IR luminosity of this source is $L_{\rm FIR}\sim
6.5\times 10^{12}$\,L$_\odot$.  More information about this galaxy is
provided by Soucail et al.\ (1999) and Kneib et al.\ (2002).

{\it SMM\,J22471$-$0206} -- There are several possible candidate
counterparts to this submm source (Fig.~6), at least two of which (P1
and P4) exhibit sufficiently unusual morphologies to be worthy of
further consideration.  The morphologies of these in the {\it HST}
$I$-band image of this field show P4 has a bright, compact core with an
unresolved arc of emission around it, while P1 exhibits a faint,
`tadpole'-like tail (Fig.~6).  Barger et al.\ (1999a) obtained optical
spectroscopy of both of these galaxies and conclude that P4 has
$z=1.16$ and shows weak AGN characteristics (broad Mg\,{\sc
ii}\,$\lambda$2800 emission), while P1 ($K=20.1$, $(I-K)=3.0$) has a
featureless spectrum and probably lies at $z\sim 2$.  The relatively
red colours and AGN signatures in P4 suggest that it is likely to be
the correct counterpart, although the low redshift of this galaxy
compared to the radio-submm spectral index estimates (Table~5) does
raise some doubts.

{\it SMM\,J02400$-$0134} -- This source was classified as a `blank
field' by Smail et al.\ (1998) and that classification still stands.
This is the brightest submm source in the sample without a reliable
counterpart, but given the depth of the available near-IR data and the
properties of some of the confirmed counterparts to brighter submm
sources, this is not particularly surprising (Table~4).  Deeper near-IR
observations of this region are required to identify any possible
counterpart to this source.  We note that there are no obvious
counterparts in either the {\it ISOCAM} 15-$\mu$m map (Metcalfe et
al.\ 1999; Biviano et al.\ 2000) or the {\it Chandra} hard X-ray image
of this field (Bautz et al.\ 2000) which both argue for a high redshift
on $K$-correction grounds.

{\it SMM\,J04431+0210} -- The near-IR counterpart, N4, to this source
is one of the EROs identified in the sample (Smail et al.\ 1999a). This
galaxy lies very close to an edge-on spiral member of the foreground
cluster lens, which was the identification originally proposed by Smail
et al.\ (1998). Its extreme optical-IR colour mean that it is invisible
on deep {\it HST} and Keck optical images of this field, although it is
well-detected in the $K$-band with an estimated FWHM of $0.4\pm 0.1''$
after correcting for lens amplification.  Smail et al.\ (1999a) discuss
various redshift constraints for this source and suggest that it
probably lies at $z\sim 3\pm 0.5$ and has $L_{\rm FIR}\sim 5\times
10^{12}$\,L$_\odot$.  There is a small additional contribution to the
amplification of this source caused by the proximity of the foreground
spiral galaxy.

\medskip
\noindent The following sources are the fainter half of our sample,
with intrinsic fluxes of $\ls 4$\,mJy.  Our follow-up of these sources
has not been as extensive as the studies of the brighter sources
discussed above, and so the information available is more limited and
our discussion is necessarily more speculative.
\smallskip

{\it SMM\,J21536+1742} -- There are several optically faint candidate
counterparts in the vicinity of this submm source.  Barger et al.\
(1999a) obtained spectroscopy of two: K2, for which they suggest a
possible redshift of $z=1.60$ from a low signal-to-noise spectrum; and
K3, an absorption-line galaxy at $z=1.02$ which appears to be weakly
detected in the {\it ISOCAM} 15-$\mu$m image of this field at a flux
of $\sim 0.2$\,mJy (L\'emonon et al.\ 1998). K3 has $K\sim 18.5$ and
$(I-K)=4.7\pm 0.1$, excluding the northern end of the
galaxy which is apparently blended with a blue point source -- the
effect of this blend on the spectroscopic observations is unknown.
The colours of K3 are too red for a passive stellar population at
$z\sim 1$ and suggest that dust plays a role in its apparent
properties, in agreement with its detection at 15\,$\mu$m.  Given the
unusual properties of K3 we suggest that it may  contribute to
the submm emission in this region, perhaps confusing an optically
fainter submm source closer to the position of K2.  The reduced
sensitivity of the 1.4-GHz map of this field (due to the bright
central galaxy) hinders the resolution of this question.

{\it SMM\,J00265+1710} -- The bright galaxies to the south of this
source (M6 in Fig.~6) are in the foreground of the cluster at $z=0.21$
(Barger et al.\ 1999a) and, apart from these, no additional
spectroscopy is available.  Smail et al.\ (2000) identify a radio
source outside of the nominal submm error box which could be the
counterpart.  This radio emission may be associated with the faint,
extended galaxy to the north-east which has $K=19.7\pm 0.2$ and
$(I-K)=2.8\pm 0.2$.  We note that 15-$\mu$m observations were taken of
this field with {\it ISOCAM}, which may help in disentangling the
identification of this source.

{\it SMM\,J22472$-$0206} -- There are several faint galaxies in the
vicinity of this submm source.  Barger et al.\ (1999a) suggested a
possible redshift of $z=2.11$ for P2, consistent with the non-detection
of this submm source at 1.4\,GHz.  To the south, P6 is relatively faint
and blue ($I=24.5$, $(I-K)<2.8$), and a very faint and apparently very
red galaxy (P7 in Fig.~6) is just visible to the south of an $I\sim 24$
galaxy east of P2. P7 is only just detected in our UKIRT $K$-band
image, but it appears to be an ERO with $K=20.7\pm 0.3$ and $(I-K)>5$.
Again this field is a candidate for a confused source arising from a
blend of two submm sources associated with P2 and possibly P7.

{\it SMM\,J00266+1710} -- Another apparently blank field.  The proximity
of the arclet, M3, to this source led Smail et al.\ (1998) to suggest
this was a possible counterpart.  However, the comparison of the arclet's
redshift, $z=0.94$ (Barger et al.\ 1999a) and the redshift estimate from
the $\alpha^{850}_{1.4}$ limit, $z>1.9$, suggests that this interpretation
is incorrect, and the true counterpart must therefore be significantly
fainter; we therefore classify this is as a probable blank field.
Nevertheless, the highly distorted morphology of the arclet, produced
by the gravitational lens, gives a graphic illustration of the likely
amplification suffered by this submm source.  Indeed, if the source
redshift is significantly greater than $z\sim 1$ it is probable that
this source actually comprises a blend of the brightest two images of
a multiply-imaged submm galaxy.  The third image would be below the
detection threshold of our SCUBA map.

{\it SMM\,J00267+1709} -- This field is classified as blank in the
analyses of Smail et al.\ (1998) and Barger et al.\ (1999a).  Our
relatively shallow UKIRT $K$-band image has not revealed any galaxies
with $K<20.6$ which are undetected in the {\it HST} $I$-band image
(Fig.~6) and so our classification for this source remains as a blank
field.

{\it SMM\,J04433+0210} -- This submm source has the lowest apparent
flux of any of the sources in our catalogue and unfortunately falls
outside the archival {\it WFPC2} F702W exposure of this field
(Smail et al.\ 1998).  Optical imaging covering this source was only
acquired after the spectroscopic survey of Barger et al.\ (1999a) was
completed. There appears to be a relatively bright candidate optical
and radio counterpart to this submm source, N5.

\medskip
\noindent The following two submm sources are both identified with the
central galaxies in the lensing clusters used in our survey:
\smallskip

{\it SMM\,J21536+1741} -- This source corresponds with the central
galaxy in the lensing cluster A\,2390 ($z=0.23$).

{\it SMM\,J14010+0252} -- This source has also been identified with
the central galaxy in the cluster lens A\,1835 at $z=0.25$. Unlike
SMM\,J21536+1741, this galaxy appears to be undergoing significant
star-formation activity (see Edge et al.\ 1999 for a more detailed
discussion of both central galaxies).

%
%
\begin{figure}
\centerline{\psfig{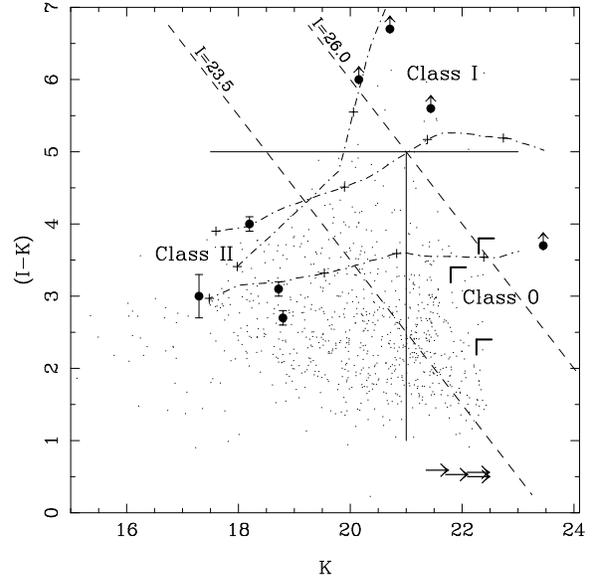} }
\noindent{\small\addtolength{\baselineskip}{-3pt}}
\caption{The distribution of proposed and confirmed counterparts to the
submm sources on the $(I-K)$--$K$ colour-magnitude plane.
Sources with possible counterparts which are only detected in the
optical passbands are marked with $\tee$. ~Sources with
no counterparts in $K$ or $I$ are plotted in the lower-right corner and
we assume the same $(I-K)$ limit as that measured in $(R-K)$ for
SMM\,J14009+0252.  The apparent magnitudes have been corrected
assuming the lens amplifications listed in Table~4.  We also plot, for
comparison, the distribution of a deep $K$-selected field sample as
small points (L.\ Cowie, priv.\ comm.) and we show the rough boundaries
of the classification scheme for counterparts to submm sources proposed
by Ivison et al.\ (2000b).  The dashed lines show the approximate
$I$-band magnitudes limits (corrected for lens amplification) for
spectroscopic identification on a 10-m telescope and deep imaging with
a ground-based 10-m or {\it HST}. The three dot-dashed lines show the
colour tracks expected for ULIRGs with $M_K=M_K^\ast+2$ and the
optical-UV SEDs measured by Trentham et al.\ (1999).  These tracks
start at $z=1$ (at $K\sim 18$) and tick marks show increments of
$\Delta z=1$. }
\end{figure}

%
%
\section{The properties of the submm population}

We begin our discussion of the nature of the submm population by
focusing on the properties of these galaxies as outlined in the
previous sections.  We start with those observables which are
available for the whole sample, such as the characteristics of their
SEDs, before turning to more detailed information which only exists
for a subset of the sample.  As the detailed follow-up work has
concentrated on the brighter sources in our survey (e.g.\ Frayer et
al.\ 2000), we will rely on their properties when discussing the
detailed characteristics of the submm population, bringing in the
fainter sources where there are suggestions of differences in their
behaviour.

\subsection{Optical and near-infrared properties}

%
%
\begin{figure*}
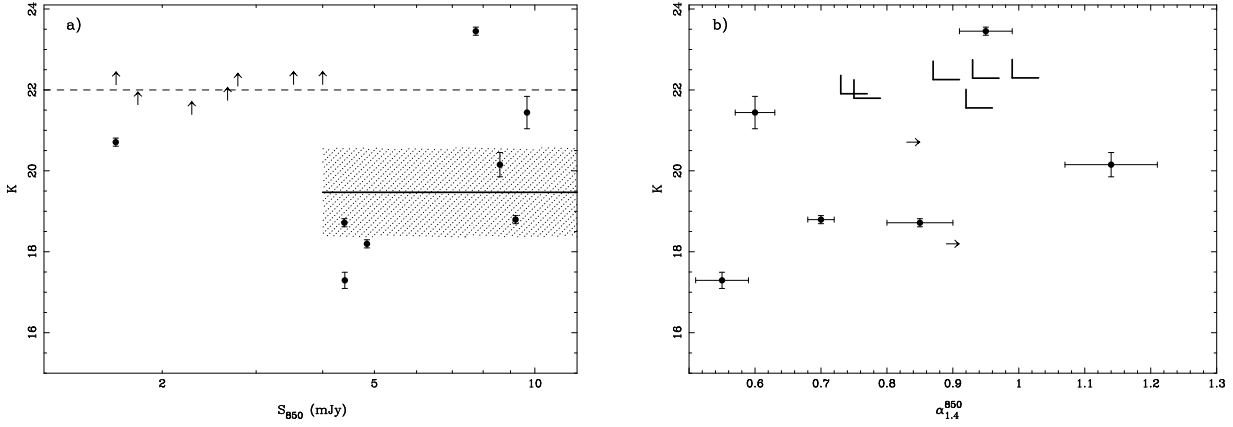

\centerline{\psfig{file=f9a.ps,angle=0,width=3.0in} \hspace*{0.3in}
\psfig{file=f9b.ps,angle=0,width=3.0in}}
\noindent{\small\addtolength{\baselineskip}{-3pt}}
\caption{a) The distribution as a function of apparent 850-$\mu$m flux
of the apparent $K$-band magnitudes of counterparts for the submm sources.
The dashed line indicates the approximate source-plane limits of our
UKIRT $K$-band imaging.  The solid line and hatched region shows the
median value and 1-$\sigma$ confidence limits for the $S_{850}\geq
4$\,mJy population.  Note the very wide range in $K$-band magnitude
exhibited by the brightest submm sources, $S_{850}\gs 4$\,mJy,
whereas the faintest submm sources do not typically possess bright
$K$-band counterparts.  b) The distribution of the $K$-band magnitude
of the counterparts against radio--submm spectral index,
$\alpha^{850}_{1.4}$.  There appears to be a weak trend for fainter
counterparts to have higher $\alpha^{850}_{1.4}$ values, although the
scatter is considerable.  The 850\,$\mu$m flux and $K$-band magnitudes
in both panels are corrected for lens amplification using the values
given in Table~4.  }
\end{figure*}

The discussion in \S5 of the optical and near-IR counterparts to the
individual submm sources highlighted the wide range in the properties
of submm galaxies at these wavelengths: although the sources span less
than an order of magnitude in their submm fluxes, their counterparts
span three orders of magnitude in their optical fluxes.  In an attempt
to identify some structure in this variety, Ivison et al.\ (2000b)
suggested a phenomenological classification scheme for the counterparts
to submm sources, analogous to that used for protostars (Adams et al.\
1987).  They defined three broad classes related to the observational
properties of likely counterparts using operational definitions based
upon the typical depths achieved in follow-up observations ($I\sim 26$,
$K\sim 21$, Fig.~8).  This magnitude limit in the near-IR corresponds
roughly to an unobscured $L^\ast$ galaxy at $z>2$--2.5, whereas the
effective dividing line in colour, $(I-K)>5$, usefully separates
the colours expected from unreddened stellar populations at $z\ls 6$
from those exhibited by dusty galaxies (Pozzetti \& Mannucci 2000).
This classification scheme therefore provides a crude differentiation
between counterparts where dust is likely to play a significant role in
the optical/IR colours and luminosity of the galaxy from those where it
does not.

In the Ivison et al.\ scheme, Class-0 objects are faint in both optical
and near-IR passband, with $K\gs 21$ and $I\gs 26$; these are likely to
be either highly obscured or very high-redshift galaxies (here we also
include those sources which are sufficiently faint in the optical that
they cannot be spectroscopically identified, $I\gs 23.5$).  Class-I
sources have counterparts in the near-IR brighter than $K\sim 21$, but
not in the optical ($I>26$) and hence will be EROs.  Finally, Class-II
sources have obvious optical and near-IR counterparts. When
spectroscopic information is available for Classes I and II, they can
be further subdivided into the  three spectroscopic types used to
classify local ULIRGs (e.g.\ Sanders et al.\ 1988): $a$, pure
starburst; $b$, type-II AGN, narrow-line but high excitation spectra
such as Seyfert 2; and $c$, type-I AGN, e.g.\ classical broad-line AGN
such as Seyfert 1. Class II submm sources may overlap with the most
strongly star-forming and massive LBGs and QSOs.

We plot the possible counterparts to the 15 submm sources discussed in
\S5 on the $(I-K)$--$K$ colour-magnitude plane in Fig.~8 and overplot
the classification boundaries.  We also show the colour--magnitude
redshift tracks expected for luminous galaxies with UV/optical SEDs
similar to the three ULIRGs studied by Trentham et al.\ (1999).  These
illustrate the wide variation in rest-frame UV/optical obscuration
encountered in this population locally.

We find that our sample consists of 8--9 Class-0 counterparts (60 per
cent), 2--3 in Class~I (15 per cent) and 4 (25 per cent) in Class~II,
depending upon the exact boundaries adopted.  It is clear that the
majority of the submm sources are extremely faint in optical wavebands
and will thus be difficult to detect in optical surveys for high-redshift
star-forming galaxies.  The four Class-II galaxies for which there are
detailed spectroscopic observations can be further sub-divided into one
Class-IIa star-forming galaxy and three Class-IIb obscured AGN.
Although many of the submm sources with optical counterparts exhibit
some level of AGN activity, the presence of strong stellar absorption
features in their optical spectra indicate that the non-thermal
emission does not dominate in the optical/near-IR wavebands.

Comparing the colours and magnitudes of the submm sources with the
tracks from Trentham et al.\ (1999) in Fig.~8, we see that the four
Class-II galaxies have colours and brightnesses consistent with those
expected for $\gg L^\ast$ galaxies with ULIRG-like UV/optical SEDs at
$z\sim 1$--3.  The 2--3 Class-I sources have colours similar to the
reddest of the local ULIRGs placed at $z\gs 2$, with apparent
magnitudes suggesting that they too have extreme rest-frame optical
luminosities, $\sim 5L_K^\ast$ (much higher than seen in similar local
systems, Trentham priv.\ comm.).  Clearly, the luminous submm galaxies
in both of these classes have already built up considerable stellar
populations. Finally, even the relatively blue colours of some of the
optical sources within the Class~0 error boxes are still consistent
with the colours of the bluest local ULIRGs, and so cannot
be ruled out as possible counterparts to the submm sources without
further information.

Figure~9a displays the $K$-band magnitudes or limits for the submm
galaxies, as a function of their apparent 850-$\mu$m fluxes (both
corrected for lens amplification).  As expected, the very different
behaviour of the $K$-correction in these two wavebands results in
little correlation between the brightness of galaxies in the submm and
near-IR wavebands, with the brightest submm galaxies in our sample
spanning a range of 6 mag.\ ($250\times$) in the $K$ band.
Nevertheless, it does appear that bright near-IR counterparts are not
seen for the faintest submm galaxies, with $S_{850}\ls 4$\,mJy, which
typically show only very faint $K$-band counterparts (the majority
simply being upper limits).  This may reflect real differences in the
mix of classes seen at high and low far-IR luminosities, or
alternatively could simply result from the small sample size.  Either
way, the possibility of a difference between the bright and faint submm
galaxies undermines attempts to extrapolate the properties of the most
luminous submm sources to the fainter sources which dominate the far-IR
extragalactic background (FIRB).

We conclude this section by briefly discussing the near-IR (rest-frame
optical) morphologies of the Class-I and II sources (Fig.~6; see also
Ivison et al.\ 1998a, 2000a and Smail et al.\ 1999a).  Where the
available imaging has sufficient signal to noise we find that all of
the sources have extended restframe optical emission (\S5), with
intrinsic FWHM of 0.5--1$''$ (or 5--10\,kpc at $z\sim 2$--3), with
several comprising multiple components (e.g.\ L1/L2, J1/J2, L3).
Concentrating on the multiple-component systems, L1/L2 and J1/J2, we
note that the spatial separation of the individual components within
these systems is 2--3$''$, corresponding to $\sim 10$\,kpc in the
source plane.  Such large separations are claimed to be relatively rare
in local ULIRG mergers (Murphy et al.\ 1996; Solomon et al.\ 1997;
although see Rigopoulou et al.\ 1999) and may suggest that ULIRG phase
occurs at a much earlier stage of mergers at high redshift, either
because of their higher gas fractions or due to other factors which
lower the stability of their gas reservoirs (e.g.\ the absence of a
strong bulge component, Mihos \& Hernquist 1996; Bekki et al.\ 1999).
A deeper survey of the restframe optical morphologies of the complete
sample must await the revival of {\it NICMOS} on-board {\it HST}.

\subsection{Submm, far-IR and radio properties}

%
%
\begin{figure}
\centerline{\psfig{file=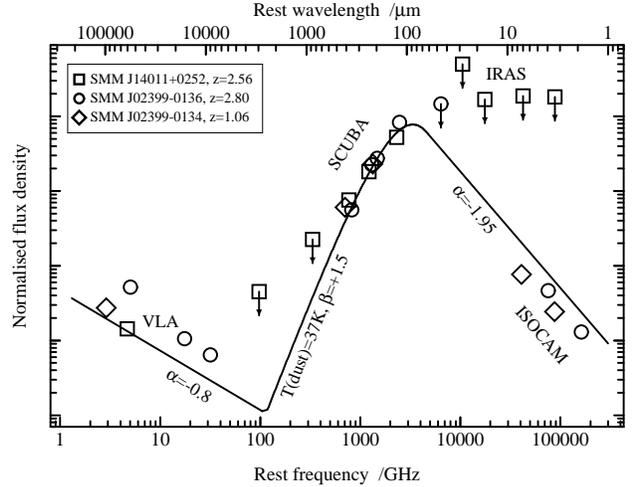,angle=0,width=3.0in} }
\noindent{\small\addtolength{\baselineskip}{-3pt}}
\caption{The rest-frame SEDs for the three submm galaxies in our survey
with reliable spectroscopic redshifts: SMM\,J02399$-$0136 (L1/L2),
SMM\,J14011+0252 (J1/J2) and SMM\,J02399$-$0134 (L3), normalised in
flux density at a rest-frame wavelength of 250\,$\mu$m. We overplot the
best-fit SED used to model the FIRB and counts, as discussed in \S6.5:
a greybody with $T_{\rm d}=37$\,{\sc k}, $\beta=+1.5$, with power-laws
of index $\alpha=-1.95$ (where $F_{\nu} \propto \nu^{+\alpha}$) beyond
60\,$\mu$m, and $\alpha=-0.80$ from 1.4\,GHz to 3\,mm.}
\end{figure}

The submm galaxies for which we have good information on their
rest-frame SEDs are naturally those with reliable redshifts.  We show
in Fig.~10 the SEDs for SMM\,J02399$-$0136 at $z=2.80$,
SMM\,J14011+0252 at $z=2.56$ and SMM\,J02399$-$0134 at $z=1.06$,
covering the rest-frame wavelength region from the near-IR to the
radio.  The spectroscopic observations show that SMM\,J02399$-$0136 and
SMM\,J02399$-$0134 both host obscured AGN, while SMM\,J14011+0252
appears to be a pure starburst (\S5). Of the three, SMM\,J02399$-$0136
is the most luminous, but the other two systems have comparable
bolometric luminosities.  However, irrespective of their different
optical spectra and luminosities, Fig.~10 shows that these three
submm-selected galaxies have very similar far-IR SEDs.  The only
significant difference occurs in the radio waveband, at rest-frame
frequencies about 4\,GHz, where the three radio-quiet galaxies differ
by a factor of $\sim 5$ in flux density. The brightest is
SMM\,J02399$-$0136, followed by SMM\,J02399$-$0134 and then
SMM\,J14011+0252.  We interpret this sequence as arising from an
increasing contribution from an active nucleus going from the pure
starburst (SMM\,J14011+0252) to a luminous type-2 AGN in
SMM\,J02399$-$0136. Apart from a weak contribution in the radio, this
change makes little overall difference to the radio--far-IR SEDs of the
galaxies, however it does illustrate the degree of uncertainty which
should be expected when trying to use the radio--submm spectral index
as a redshift estimator for galaxies hosting radio-loud AGN. The range
in rest-frame radio power shown by the three galaxies compared here
would result in a systematic reduction of the spectral index by
$\delta\alpha^{850}_{1.4}=-0.3$, and an equivalent reduction in the
estimated redshift of the source.

At the shorter wavelengths covered by Fig.~10, we see that the SEDs
are poorly constrained between 10--100\,$\mu$m in the rest frame, a
situation that should soon be resolved by SOFIA and {\it SIRTF}. At
the slightly longer wavelengths observed by SCUBA, we find the average
SED of the three galaxies is consistent with the characteristic
spectral index $\alpha \simeq 3.5\pm0.5$, of optically-thin emission
from dust grains.  Simple fits to this composite SED cannot, of
course, accurately constrain the dust temperature, although the data
for all three galaxies are consistent with the $T_{\rm d}$ range of
35--50\,{\sc k} found for other dusty, high-redshift systems (Benford
et al.\ 1999).  Higher temperatures can only be fitted if the opacity
of the dust becomes significant at wavelengths of about
100--200\,$\mu$m.

As mentioned in \S4.6, the redshifts estimated from the spectral
index are sensitive to the assumed temperature of the dust, $T_{\rm
d}$, in the submm galaxies (Blain 1999) and so reducing $T_{\rm d}$
would allow a lower median redshift for the population.  However,
only if we force the entire submm population to have a characteristic
dust temperature less than 30\,{\sc k} can we start to push the median
redshift much below $<\!  z\! >\sim 2$.  There is no evidence for such
a cold characteristic temperature in any of the brighter submm galaxies
with confirmed identifications and redshifts, nor is there any physical
reason to expect that dust in high-redshift starbursts should be any
cooler than that in local LIRGs and ULIRGs.

Looking at the population as a whole, if typical submm-selected
galaxies have cool dust temperatures (and thus lay at low redshifts),
for example an analogue  of the Milky Way  with $\sim 17$\,{\sc k}
(Reach et al.\ 1995), they would be detectable at the mJy level using
SCUBA and at the $\mu$Jy level using the VLA, consistent with the
observations.  However, their optical counterparts would be very bright
in {\it HST} and ground-based follow-up images. Based on the data
presented here, such sources can only represent a few per cent of
SCUBA-selected galaxies.

Is there any evidence for differences in dust temperature within the
submm population?  Fig.~9b illustrates the variation in
amplification-corrected apparent $K$-band magnitude with radio-submm
spectral index of the source.  There may be a weak tendency for sources
with higher $\alpha^{850}_{1.4}$ indices to have fainter counterparts,
but any trend is weakened by the large scatter, a factor of 100
at a fixed $\alpha^{850}_{1.4}$.  This large scatter may reflect both
the crudeness of $\alpha^{850}_{1.4}$ as a redshift indicator and the
strong variation in the rest-frame UV/optical obscuration in the submm
population.

Higher values of $\alpha^{850}_{1.4}$ are expected to correspond to
galaxies with either higher redshifts for the same SED, or to cooler
dust temperatures for the same redshift (Blain 1999). The observations
are consistent with either explanation. If the redshift distribution is
systematically higher for fainter sources -- a straightforward
explanation, and one that is hinted at in Fig.~9b -- then it will be
difficult to test this idea, as the brightness of the counterpart to the
submm source will be reduced in all other wavebands.  The distribution
of the $K$-band counterparts in Fig.~9a suggests that the fainter submm
sources may indeed have fainter near-IR counterparts, which would
support this suggestion, although this could simply indicate that the
fainter sources are more obscured. Alternatively, if the fainter
sources are systematically cooler, then the bolometric luminosity
derived for them, as associated with a given 850-$\mu$m flux density,
will be much less as compared with that estimated using a hotter dust
temperature, and so their counterparts in other wavebands would also be
expected to be systematically fainter; however, the typically lower
redshift of the counterparts might make their detection in the $K$-band
easier in this case. The resolution of this issue awaits both accurate
multi-band photometry of submm-selected galaxies using {\it SIRTF} and
SOFIA at wavelengths corresponding to the rest-frame peak of their SED,
and the measurement of redshifts for submm galaxies over a wider range
of flux density.

Using the new estimates for the redshifts and limits from \S4.6.4, listed
in Table~5, we conclude that at least half of the submm population down
to a median {\it intrinsic} 850-$\mu$m flux of $\sim 3$--4\,mJy have
redshifts above $z=2$.  This result is consistent with the expectations
from the 450-/850-$\mu$m flux ratios discussed in \S4.6.3.

If the median redshift of the submm population is $z>2$ and we adopt a
mid-IR SED for the galaxies similar to Arp\,220 then their predicted
15-$\mu$m fluxes would typically be $<10\mu$Jy -- well below the {\it
ISOCAM} detection limit.  The relatively high rate of {\it ISOCAM}
15-$\mu$m detections we find therefore suggests that Arp\,220 is not a
good prototype for the mid-IR SED of a faint submm galaxy, which are
typically brighter at mid-IR wavelengths.  Fig.~11 illustrates the
mid-IR--submm spectral indices of the current sample of galaxies with
both {\it SCUBA} and {\it ISO} observations (or limits) along with the
behaviour expected from a number of SEDs.  The observations appear to
rule out Arp\,220--like SEDs in the mid-IR (which are too red), but are
well described by model SEDs of less optically-thick starbursts (Dale
et al.\ 2001; Eales et al.\ 2000).  Fig.~11 also illustrates the
behaviour of the mid-IR--submm spectral index from the composite SED
used by Blain et al.\ (1999b) to model the FIRB and counts (see \S7),
including the faint counts of galaxies detected by {\it ISOCAM} at
15\,$\mu$m and the deep counts of radio galaxies determined at
frequencies of 1.4 and 8.4\,GHz using the VLA (see \S6.5).  This SED
has $T_{\rm d}=37$\,{\sc k}, $\beta=+1.5$ and a mid-IR spectral index
of $-1.95$, and lies on the low-redshift far-IR--radio correlation. It
clearly provides an adequate representation of the behaviour of
individual galaxies in the mid-IR waveband.

%
%
\begin{figure}
\centerline{\psfig{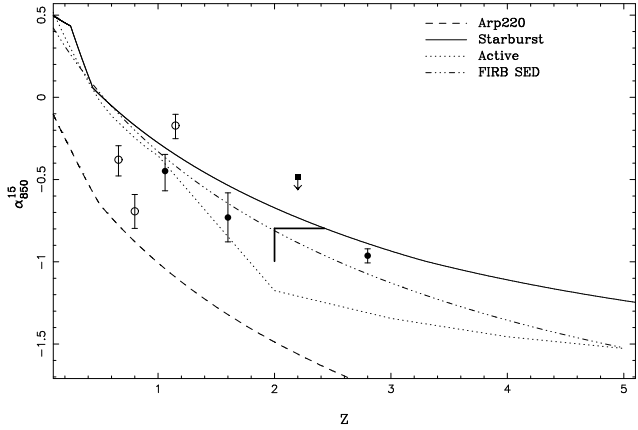} }
\noindent{\small\addtolength{\baselineskip}{-3pt}}
\caption{The variation in the spectral index between 15 and
850\,$\mu$m as a function of redshift for a range of SEDs.  We show
the expected variation in the spectral index for the SED of
Arp\,220, the starburst SED from Eales et al.\ (2000), the most active
SED from Dale et al.\ (2001) and the best-fit SED used to fit the mid-
and far-IR background from Blain et al.\ (1999b) (assuming $T_{\rm
d}=37$\,{\sc k}, $\beta=+1.5$ and a mid-IR spectral index, $-1.95$).
We plot the measurements from the SCUBA Lens Survey as filled symbols,
the open symbols show joint SCUBA and {\it ISOCAM} detections from
Hughes et al.\ (1998) and Eales et al.\ (2000).  The upper limit
represents the typical limits on the $\sim$80 per cent of the submm
sources undetected by {\it ISOCAM} from Eales et al.\ (2000) and
Hughes et al.\ (1998).  }
\end{figure}

\subsection{Dynamical, gas and dust masses}

Dynamical information is available for three galaxies in our sample:
SMM\,J02399$-$0136 (Frayer et al.\ 1998); SMM\,J14011+0252 (Frayer et al.\
1999; Ivison et al.\ 2001b) and SMM\,J02399$-$0134 (Kneib et al.\ 2002).
This comes from mapping redshifted CO line emission in these galaxies
with the OVRO and IRAM interferometers.  The width of the molecular
line provides an estimate of the dynamical mass of the system, while
the strength of the line can be used to estimate the total molecular
gas mass of the systems, assuming a conversion from CO gas mass to total
molecular gas mass (mostly H$_2$), an assumption which may be uncertain
at the factor of $\gs 4$ level (Downes \& Solomon 1998; Papadopoulos et
al.\ 2001).

It is interesting to note that although SMM\,J02399$-$0136 and
SMM\,J14011+0252 have very different optical classifications, they have
very similar (and high) far-IR to CO ratios:  $L_{\rm FIR}/L'_{\rm
CO}\sim 500$\,L$_\odot$\,{\sc k}$^{-1}$\,km\,s$^{-1}$\,pc$^{-2}$, at the
high end of the values seen in local ULIRGs (Solomon et al.\ 1997).
This suggests that their prodigious far-IR luminosities are powered by
similar physical processes, probably highly efficient star formation
(Solomon 2001).

The dynamical masses in the central 10--20\,kpc infered for these
systems are 1--$2\times 10^{11} \sin^{-2}(i)$\,M$_\odot$ (where $i$
is the unknown inclination angle), while their total molecular gas
masses are estimated at 1--$2\times 10^{11}$\,M$_\odot$, assuming a CO
luminosity to molecular gas mass conversion factor, $\alpha=4$\,M$_\odot$
({\sc k}\,km\,s$^{-1}$\,pc$^2$)$^{-1}$.  This suggests that molecular
gas is a dynamically important, and perhaps dominant, constituent in the
central $\sim 10$\,kpc of these galaxies (see Frayer et al.\ 1998, 1999).
Combining these dynamic mass estimates with the measured rest-frame
optical luminosities of these galaxies we find very low mass-to-light
ratios: M$_{\rm dyn}$/L$_V\sim 0.3$, as expected for star-forming systems
dominated by very recent star formation, $\sim 100$\,Myrs (Salasnich et
al.\ 2000).

Adopting $k_{\rm d} = 0.15 (\lambda_0/800\,\mu{\rm m})^{-1.5}$ for the
standard dust emission parameter, where $\lambda_0$ is the rest-frame
wavelength, in order to compare the results with other distant sources,
we estimate dust masses of $M_{\rm d}= 0.5$--$5 \times
10^8$\,M$_{\odot}$ for the typical submm galaxies in our sample, based
on dust temperatures of $T_{\rm d} = 40\pm 10$\,{\sc k} (see \S6.2).
Given the modest dust yield from a standard IMF, these large dust
masses provide further support for the high SFRs claimed for these
galaxies, assuming of course that type-II supernovae are a significant
source of dust in the early Universe, which has yet to be demonstrated
convincingly.

\subsection{Power source: AGN versus starburst}

The source of the extreme luminosities of the submm galaxy population
detected by SCUBA remains a key problem for interpreting their nature
and their relevance for models of galaxy formation and evolution.
Unfortunately, submm observations alone are unable to discriminate
between luminous dusty galaxies powered by massive stars or AGN,
although the resolution of a handful of sources in the radio or
millimeter on scales of $\ls 5''$ suggests that the submm emission from
these galaxies comes from an extended region rather than a central AGN
(or in fact nuclear starburst).  This result cautions against the use
of very high resolution radio or millimeter facilities for surveys to
identify this population.

While it is plausible that AGN powered dusty galaxies could on average
have hotter dust temperatures, rendering them less likely to be selected
in submm surveys (Fig.~1), at present there is insufficient information
to confirm this.  We therefore have to resort to secondary indicators
of AGN activity to determine their prevalence in the submm population
as a whole.  We stress that even when an AGN is known to be present,
determining its contribution to the overall energetics of the submm
emission is still far from trivial (e.g.\ Frayer et al.\ 1998, 1999).

There are at least three unambigious examples in our sample of submm
galaxies which harbour AGN nuclei: L1/L2, L3 and P4 (Table~5), all
based on broad lines visible in optical spectroscopy.  In all three
cases these AGN show signatures of dust obscuration in their line
ratios.  For the optically-faint majority of the sample, however, we
have to rely on other tracers of AGN activity, the most powerful
being hard X-ray emission (Fabian et al.\ 2000; Bautz et al.\ 2000;
Hornschemeier et al.\ 2000; Page et al.\ 2001).

Using the {\it ROSAT} X-ray limits from \S4.5 and our observed submm
fluxes for the individual sources, we estimate that the typical submm
galaxy has a submm--X-ray spectral index, $\alpha^S_X$ (Fabian et al.\
2000), of $\alpha^S_X\gs 1.1$ (assuming a $\Gamma=2$ power-law for the
X-ray emission).  We compare the limits on $\alpha^S_X$ for the submm
sources with five quasars at $z>4$ using the combined {\it ROSAT} X-ray
observations of Kaspi et al.\ (2000) and submm photometry from McMahon et
al.\ (1999).  These quasars have $\alpha^S_X\sim 0.9\pm 0.1$, with the
most submm-luminous example being BR\,1202$-$0725 with $\alpha^S_X\sim
1.1$.  We conclude that if AGN are present in typical submm galaxies
then they are more obscured than the dusty and gas-rich $z=4.69$ QSO
BR\,1202$-$0725 (Omont et al.\ 1996).  To better quantify this statement,
we use the obscured AGN models from Fabian et al.\ (2000) and Gunn \&
Shanks (2001) and suggest that if AGN are present in the typical submm
galaxy then they must be lie behind columns of $N({\rm H}{\sc i}) >
10^{23}$\,cm$^{-2}$.  The more sensitive {\it Chandra} observations of a
smaller sample suggest that if submm galaxies host AGN then these must
be Compton-thick ($N({\rm H}{\sc i}) \gg 10^{24}$\,cm$^{-2}$) and the
amount of unreprocessed AGN emission which does escape by scattering must
be small (Fabian et al.\ 2000; Bautz et al.\ 2000; Hornschemeier et al.\
2000; Almaini et al.\ 2001).

Thus the available information suggests that the fraction of SCUBA
galaxies which host optically-identifiable AGN is between 10 and 20 per
cent (Barger et al.\ 1999a) and that if AGN are present in the majority
of the submm population then these must be sufficiently obscured that
they are unlikely to be visible in the optical anyway.  The low
fraction of AGN-dominated systems is consistent with much more detailed
surveys of similar luminosity ULIRGs at $z\sim 0$, which find 20--30
per cent are AGN-dominated (Genzel et al.\ 1998).

Better constraints on the proportion of AGN in the submm population
await more sensitive hard X-ray observations with {\it XMM/Newton} (in
blank fields) or {\it Chandra} (Almaini et al.\ 2001), searches for
broad emission lines in the less dust-sensitive near-IR wavebands and
surveys for hot dust emission from AGN-illuminated tori in the mid-IR.

\subsection{Constraining the evolution of dusty galaxies}

Finally, we discuss the exploitation of the submm population
for understanding the wider issue of the evolution of dusty galaxies.
This section revisits the results of Blain et al.\ (1999b,1999c) to
constrain simple models for the evolution of luminous dusty galaxies
and provide some physical insights into the processes driving their
activity.

In the local Universe, the luminosity function of luminous dusty galaxies
is best constrained at 60\,$\mu$m based on the {\it IRAS} survey (Saunders
et al.\ 1990).  Additional information about these galaxies is also
available at 100\,$\mu$m (Soifer \& Neugebauer 1990).  These wavelengths
are close to the peak of the SED for any reasonable dust temperature
and the ratio of the bright counts at 60 and 100\,$\mu$m implies a
luminosity-averaged dust temperature $T_{\rm d} \sim 35$--45\,{\sc k}.
850-$\mu$m observations of galaxies detected by {\it IRAS} (Dunne et al.\
2000; Lisenfeld et al.\ 2000) provide a longer wavelength baseline and
hence an excellent probe of the SED.  These indicate $T_{\rm d} = 36
\pm 5$\,{\sc k} and a Rayleigh--Jeans spectral index of $3.3 \pm 0.2$
(i.e.\ $\beta=+1.3$, but see Dunne \& Eales 2001). The population of
low-redshift dusty galaxies can be divided into relatively short-lived
warm interacting/starbursting galaxies and long-lived cooler quiescent
galaxies (Blain et al.\ 1999c; Barnard \& Blain 2001); however, the
details of this distinction are relatively unimportant for studies of
high-redshift galaxy evolution. Any low-luminosity, low-temperature dusty
galaxies missing from existing surveys do not contribute significantly
to the luminosity density, even at low and moderate redshifts (Chapman
et al.\ 2002).

The form of evolution of the baseline low-redshift far-IR luminosity
function $\Phi_0(L)$ must be dominated by pure-luminosity evolution,
that is $\Phi(L,z) \simeq \Phi_0[L/g(z),0]$, to ensure that the far-IR
background radiation intensity is not exceeded.  Number-density evolution
is also likely to be involved, but must be dominated by luminosity
evolution (Blain et al.\ 1999b; Chapman et al.\ 2001c).  The evolution
function $g(z)$ is determined by demanding that the background radiation
intensity, counts and redshift distributions of dusty galaxies are
all in agreement with observations. These observations are, in order
of increasing redshift, the faintest counts and redshift distributions
of 60-$\mu$m {\it IRAS} galaxies, deep 90- and 170-$\mu$m counts from
{\it ISO}, the spectrum of background radiation from {\it COBE}, and
the faint counts and limited redshift information of distant galaxies
detected using SCUBA at 450 and 850\,$\mu$m and MAMBO at 1.2\,mm.

Several approaches can be taken to investigate the evolution. The
simplest is to assume a parametric form for $g(z)$ and fit this to the
data to construct a model which relates observations across a range of
wavebands in a simple phenomenological manner (Blain et al.\ 1999b; see
also Rowan-Robinson 2001).  This has the advantage of requiring few
parameters to model the galaxy SED so that the form of evolution can be
well constrained using the sparse observational information. A more
physically motivated approach connects the evolving mass function of
galaxies to the associated luminosity function using a prescription for
both star formation and the fueling of AGN (Baugh et al.\ 1998;
Guiderdoni et al.\ 1998; Blain et al.\ 1999c); however, care must be
taken to avoid getting lost in the space of available free parameters.
Without an additional population of short-lived, very luminous
galaxies, the simplest standard semi-analytical models, which include
star formation in the gas that cools in galaxy disks, fail to account
for the observed surface density of SCUBA and MAMBO sources (Guiderdoni
et al.\ 1998).

\subsubsection{A simple parametric model}

This approach was adopted by Blain et al.\ (1999b), who assumed a
local 60-$\mu$m luminosity function, an SED defined by a single dust
temperature $T_{\rm d}$ and a form of evolution $g(z) = (1+z)^\gamma$
at low redshifts. $T_{\rm d}$ and $\gamma$ are determined from the
60-$\mu$m {\it IRAS} counts and the latest {\it ISO} counts at 90- and
175\,$\mu$m (Juvela et al.\ 2000).  We derive $T_{\rm d} = 37 \pm
3$\,{\sc k} and $\gamma = 4.05 \pm 0.15$, assuming a Rayleigh--Jeans
spectral index of +3.5.  These results differ only slightly from the
original analysis of Blain et al.\ (1999b), based on more limited {\it
ISO} data, $T_{\rm d} = 38 \pm 4$\,{\sc k} and $\gamma = 3.9 \pm 0.2$.
Equally, changing to the more popular non-zero-$\Lambda$ cosmology
does not significantly alter this result (Blain 2001).

The characteristic dust temperature we derive is consistent with
subsequent SCUBA measurements of local luminous {\it IRAS} galaxies
(Dunne et al.\ 2000), suggesting a continuity in the properties of this
population between low and moderate redshifts.  The value of $\gamma$
we find is closer to that claimed for the evolution of the optical
luminosity density (Lilly et al.\ 1996) and mid-IR counts (Xu 2000)
than the value of $\gamma = 3$ often assumed to describe the evolution
of galaxies in the far-IR waveband.

At higher redshifts, the behaviour is not well constrained by {\it ISO}
observations (although future {\it SIRTF} surveys will make a major
impact in this area) and we rely instead on the form of the background
radiation intensity (Fixsen et al.\ 1998) and the counts of SCUBA
galaxies. These are somewhat degenerate, although the SCUBA data
provide the better constraint.  The most useful of the various simple
parametric forms adopted by Blain et al.\ (1999b) to describe the
evolution of $\Phi$ at high redshifts is the so-called `Gaussian'
model, in which $g(z)$ at moderate and high redshifts is represented by
a Gaussian in cosmic epoch.  Since then, progress has been made in
developing a more appropriate form of $g(z)$, which is fully compatible
with models of cosmic chemical evolution and naturally includes a peak
in the evolution function (Jameson 1999):
\begin{equation}
g(z) = (1+z)^{3/2} {\rm sech}^2[ b \, {\rm ln}(1+z) - c ] \, {\rm cosh}^2c.
\end{equation}
At low redshifts, $\gamma \simeq (3/2) + 2b \sqrt{1-{\rm sech}^2 c}$.
Using all available observational data, the results $b=2.2 \pm 0.1$ and
$c=1.84 \pm 0.1$ are obtained; see the solid line in Fig.~12. These
results are similar to those derived by Blain et al.\ (1999b), but use
the more accurate 175- and 850-$\mu$m counts now available, as
well as the limits on the probable redshift distribution of the
850-$\mu$m population from \S4.6.

%
%
\begin{figure}
\centerline{\psfig{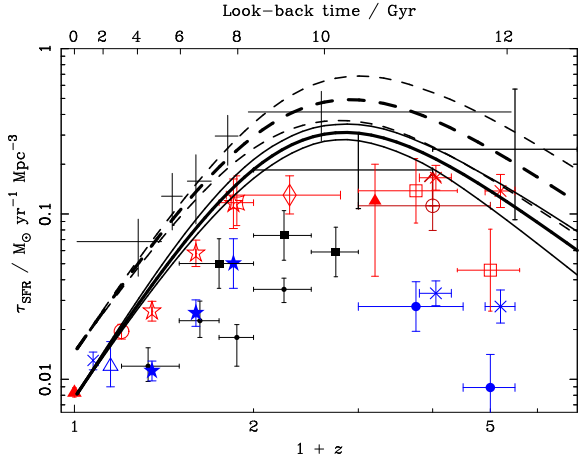} }
\noindent{\small\addtolength{\baselineskip}{-3pt}}
\caption{The history of star formation inferred using the methods
discussed in \S6.5.  The thick solid line is derived using a simple
parametric luminosity evolution model and the thick dashed line with a
hierarchical model. Both have been revised since Blain et al.\ (1999b,
1999c) were published, to reflect the increased amount of far-IR and
submm data now available.  The thin lines show the envelope of
68-per-cent uncertainty for each of the two models.  The relative
vertical normalisation of the curves depends on the assumed IMF and the
AGN fraction.  The data points are inferred from UV/optical/near-IR
observations, in order of increasing redshift, by Gallego et
al.\ (1996; filled triangle), Gronwall (1999; thin diagonal cross),
Treyer et al.\ (1998; open triangle), Tresse \& Maddox (1998; empty
circle), Lilly et~al.\ (1996; filled stars), Cowie, Songaila \& Barger
(1999; small filled circles), Glazebrook et al.\ (1999; bent square at
$z=0.9$), Connolly et al.\ (1997; filled squares), Moorwood et
al.\ (2000, filled triangle), Yan et al.\ (1999; empty lozenge), Madau
et al.\ (1996; large filled circles), and Pettini et al.\ (1998; empty
squares). Flores et al.\ (1999; empty stars) and Pettini et al.\ (1998)
have corrected the Lilly et al.\ and Madau et al.\ results respectively
for observed dust extinction.  The high-redshift points are derived
from analyses of the {\it HDF}.  Wide-area ground-based surveys by
Steidel et al.\ (1999) have increased the estimated high-redshift SFR,
as shown by the thick diagonal crosses. No extinction correction is
applied to these points; Steidel et al.\ (1999) estimate that the
extinction-corrected SFR is greater by a factor of about 5; the effects
are shown by the higher pair of diagonal crosses.  At longer
wavelengths, SFRs have been derived from mid-IR observations by Flores
et al.\ (1999) and in the submm by Hughes et al.\ (1998; empty circle
with upward pointing arrow).  Based on radio observation, supplemented
by redshift information provided by spectroscopic and photometric
redshifts, Haarsma et al.\ (2000) infer star-formation rates shown by
the thin solid crosses with no terminators. Based on a similar
technique, supplemented further by submm observations, and including a
completeness correction factor of 11, Barger, Cowie \& Richards (2000)
derive star-formation rates shown by the thick solid crosses with
terminators.  }
\end{figure}

\subsubsection{A model of merging galaxies}

Rather than use a purely parametric form for the evolution of
$\Phi(L,z)$, an alternative approach is to adopt a more
physically-motivated model based upon the merger of dark matter halos
in the Press--Schecter formalism (Press \& Schecter 1974).  This
implicitly assumes that the activity observed in the far-IR is
triggered through interactions and mergers (see \S6.1).  We use a
simple form of the evolution of the merger rate of dark-matter halos
(Blain \& Longair 1993; Blain et al.\ 1999c), which adequately
reproduces the results of recent $N$-body simulations (Jenkins et al.\
2001), and make the most basic assumption that a certain
redshift-dependent fraction $x(z)$ of the total mass of dark and
baryonic matter involved in these mergers is converted into energy by
nucleosynthesis in high-mass stars with an efficiency 0.007$c^2$.  The
same formalism is appropriate for describing the evolution of AGN
fueling events at the epochs of mergers (Blain et al.\ 1999c).  The
form of evolution and normalization of $x(z)$ can be determined in the
same manner as for the parametric model, using the background radiation
intensity and low-redshift {\it IRAS} counts.  Using an appropriate
form of $x(z) = g(z) / (1+z)^{3/2}$ (Jameson 1999), we find best-fit
parameters of $b = 1.95 \pm 0.1$, $c = 1.6 \pm 0.1$ and $x(0) = 1.35
\times 10^{-4}$ in our standard cosmology, assuming the same galaxy SED
used in the parametric model.  However, this model under-predicts the
observed 175- and 850-$\mu$m counts by a large factor and so we are
forced to introduce an additional parameter to allow the efficiency of
merger-induced bursts of activity to increase at high redshifts so as
to reproduce these long-wavelength source counts.  This additional
parameter is the product of the fraction $F$ of mergers which generate
luminous bursts and the duration of these bursts, $\sigma$. The product
$F\sigma(z)$ is included in our model and then constrained using all
the available far-IR counts, backgrounds and redshift distributions.
Using the form $F\sigma(z) = F\sigma(0) \exp(az + bz^2 + cz^3)$, we
derive values of $F\sigma(0) = 2.4$\,Gyr, $a=-4.14$, $b=-0.56$ and
$c=0.46$. The resulting history of galaxy evolution is shown by the
thick dashed line in Fig.~12.

\subsubsection{The star-formation history}

Both the parametric and hierarchical models account for all the
observed background radiation, counts and redshift distributions of
galaxies at wavelengths longer than 60\,$\mu$m.  As shown by the
comparison of the models with the optical/UV derived data points in
Fig.~12, the dominant source of energy in the Universe is the
rest-frame far-IR radiation of starlight and AGN emission which has
been reprocessed by dust.  Indeed, apart from the relative
normalisation (which can be varied by altering the assumed IMF and the
fraction of AGN-powered activity), the parametric and hierarchical
models in Fig.~12 show very similar behaviour.   Equally the
star-formation histories derived from deep radio selected samples also
show a comparable behaviour (Cram 1998; Barger, Cowie \& Richards 2000; Haarsma
et al.\ 2000; Chapman et al.\ 2001), which also tend to lie above the
optically-derived values, in broad agreement with the dust-enshrouded
star-formation histories derived here  (see Fig.\,12).  Note that the
results of Barger et al.\ (2000) do not clearly indicate a decline in
the high-redshift evolution of the star-formation rate.

In our hierarchical model, the strong increase in the far-IR luminosity
density out to $z\sim 1$--2 arises in part from the increased rate of
mergers in the early Universe. In addition, however, the ratio of the
total amount of energy released during mergers and the mass of dark
matter involved must increase sharply with redshift.  This suggests
that the progenitors of the submm galaxies which dominate the far-IR
emission at high redshifts must be more prone to massive starbursts or
AGN activity.  This conclusion echoes the discussion of individual
submm galaxies earlier in this section, where we find evidence that the
typical separations between the components in these massive, gas-rich
mergers maybe larger than has been claimed for similarly luminous galaxies
locally.  This also hints at an increased instability in the progenitor
galaxies at high redshifts as the bolometric emission from massive,
high-redshift mergers appears to peak earlier during their
interaction.  Finally, we note that to avoid biasing the redshift
distribution of SCUBA-selected galaxies to very high redshifts (Blain
et al.\ 1999b) and to remain consistent with the spectral index of the
mm-wave extragalactic background radiation (Gispert, Lagache \& Puget
2000), the comoving star-formation rate density in the obscured
population must decline beyond $z\gs 3$--4.

The framework incorporated in our models can be extended to include
(or predict) observations in the radio and mid-IR regimes.  For
example, assuming the standard form of the far-IR--radio correlation,
with a radio spectral index of $-0.65$, then the parametric and
hierarchical models predict source surface densities brighter than
10\,$\mu$Jy at 8.4\,GHz of 1.05 and 0.98\,arcmin$^{-2}$, with slopes
of $-1.4$ and $-1.3$ respectively, closely matching the observed count
$N(\ge S)=(1.01 \pm 0.14) (S/10\mu{\rm Jy})^{-1.25 \pm 0.2}$
(Partridge et al.\ 1997).  Whereas, adopting a simple form for the
mid-IR region of our fiducial galaxy SED, $F_\nu \propto \nu^{+\alpha}$
with $\alpha = -1.95$ at $\lambda\ls 60\mu$m, then we reproduce the
normalization and general features of the deep 15-$\mu$m galaxy counts
from {\it ISOCAM} (Elbaz et al.\ 1999), including the marked change of
slope at flux densities between 0.5 and 1\,mJy.  The predicted slope
of the 15-$\mu$m counts at flux densities between 1 and 10\,mJy is
steeper in the hierarchical model and provides slightly better
agreement with the observations than the parametric scheme.

Finally, we note one outstanding problem with the star-formation histories
derived from the far-IR, first highlighted by Blain et al.\ (1999b).
This is that the predicted mass fraction in stars, $\Omega_\ast$,
at the present day, from these models, exceed the estimates of the
observed stellar density by a factor of about five, based on a large
near-IR-selected redshift survey by Cole et al.\ (2001).  One possible
explanation for this discrepency is that a large fraction of the
luminosity of distant dusty galaxies could be produced by AGN rather than
by high-mass stars.  However, the current estimates of the fraction of
AGN-dominated submm sources presented in \S6.4 suggest this is not likely
(although this obviously requires confirmation, e.g.\ Ivison et al.\ 1998,
2000a; Almaini et al.\ 2001).  This indicates, therefore, that if
the background radiation intensity and counts determined in the submm
waveband are not significantly in error then either a larger fraction
of the baryons in the Universe has been processed into stars than is
estimated from local galaxy surveys or, more likely, the process of star
formation in the submm population must differ from that in nearby `normal'
galaxies, with an initial IMF biased to high-mass stars (more akin to
the that seen in 30~Doradus, Sirianni et al.\ 2001).  Blain et al.\
(1999b) estimated that a lower mass limit of about 1\,M$_\odot$
is required in the IMF in the submm galaxies to comply with the local
stellar mass fraction, although higher mass cutoffs are also feasible.

%
%
\section{Discussion}

There are several questions about the submm population which need to
be urgently addressed.  These concern the nature of the galaxies
selected in the submm, their redshifts, masses and power sources, as
well as more general issues of their relation to other classes of
high-redshift sources, such as LBGs, radio galaxies and QSOs.

As this paper has demonstrated, we are still at a relatively early
stage in our understanding of the mJy submm population.  In part, this
is because they were only discovered four years ago. However, as we
have stressed here, the main barrier to improving our knowledge of this
population is the lack of bright optical (or even near-IR) counterparts
for the bulk of the submm galaxies found to date.  As a result, we are
left with relatively little evidence to sift when trying to understand
the nature and evolution of this population and must fall back on
circumstantial evidence and some degree of conjecture when attempting
to place these galaxies within their correct cosmological context.

Integrating the deepest submm counts, we have shown that the submm
galaxies lying in the decade of flux brighter than about 1\,mJy produce
about 60 per cent of the far-IR background at 850\,$\mu$m (see also
Hughes et al.\ 1998; Eales et al.\ 1999) and probably closer to 80 per
cent by 0.5\,mJy.  Taken in combination with the fact that over half
the energy density in the extragalactic background is contained in the
far-IR and submm wavebands, it is clear that this population is
important for understanding the energetics of galaxy formation and
evolution.

The weight of evidence on the distances to these galaxies, from
optical and near-IR spectroscopy and crude limits based on far-IR and
radio colours, appears to point to the bulk of this population lying
at high redshifts, $z>2$.  This confirms that for any reasonable
characteristic dust temperature these dusty galaxies must have high
bolometric luminosities, $\gs 10^{12}$\,L$_\odot$, and thus classifies
them as ULIRGs.  The importance of this population at high redshifts
is in stark contrast with its negligible contribution to their
luminosity density at $z=0$.  Thus we must be seeing rapid evolution
in these highly luminous systems.

On the question of what powers the immense luminosities of the submm
population, again the balance of evidence is that majority of the
population do not appear to host unobscured, or partially-obscured, AGN
(e.g.\ Almaini et al.\ 2001), similar to the proportion of
AGN-dominated ULIRGs seen locally (Genzel et al.\ 1998). Even in the few
examples which do show clear signatures of non-thermal emission it
appears that star formation still contributes a large fraction of the
total bolometric output (Frayer et al.\ 1998; Bautz et al.\ 2000).

This then leaves us with massive star formation as the most likely
mechanism to power these sources.  The SFR required to produce $L_{\rm
FIR} \sim 10^{12}$\,L$_\odot$ corresponds to a formation rate of O, B
and A stars of about 210\,M$_{\odot}$\,yr$^{-1}$ (or an upper limit of
650\,M$_{\odot}$\,yr$^{-1}$, based on an extrapolation using a Salpeter
IMF extending from 0.1 to 100\,M$_{\odot}$, Thronson \& Telesco 1986).
These estimates are consistent with other measures of the SFR available
for a handful of the submm population (Ivison et al.\ 2000a) and are
sufficient to form an $L^\ast$-galaxy in only a few 100\,Myr.

Next, we address the question of the relationship between the submm
galaxies and other classes of high-redshift sources and the
information these provide about galaxy formation as traced by the
star-formation history of the Universe.

In a recent paper, Adelberger \& Steidel (2000, AS) presented a useful
analysis of the properties that  submm galaxies would be expected assuming
that they can be described by an extrapolation of the behaviour of the
populations selected in the rest-frame UV.  Based upon the rest-frame UV
emission (observed $R$/$I$ bands) and a model of the correlation between
UV spectral slope and far-IR emission for local UV-detected starbursts,
AS predict that the median magnitude of the counterparts to the 850-$\mu$m
sources in our survey should be $I\sim 24$.  However, as Table~4 shows,
the lensing-corrected median apparent magnitude for our sample is actually
$I\geq 26$, and may be considerably fainter as most of the sources
remain undetected in $I$, even though some are seen in the $K$-band.
This indicates that the UV/far-IR ratio of these submm-selected galaxies
is roughly an order of magnitude lower than expected from AS's model.

Indeed, recent work by van der Werf et al.\ (2001a), Baker et
al.\ (2001), Sanders (2001) and Meurer \& Seibert (2001) has shown that
the extrapolation used by AS is not followed by bolometrically-selected
samples of very luminous local and distant galaxies. Hence, the basic
assumption of AS's analysis -- that the SFRs of luminous submm galaxies
can be predicted reliably from their rest-frame UV emission -- is
unfounded (also see Chapman et al.\ 2000; Peacock et al.\ 2000).
Moreover, given their extreme faintness in the optical, a large
proportion of the submm galaxies are completely missed by UV-selected
surveys of the distant Universe.  Obviously any attempt to apply a
correction factor for the star formation density missed in this
population cannot rely on their (unmeasurable) UV spectral slopes.  As
a result, such surveys provide a lower limit to, but a potentially
seriously incomplete measure of, the total star formation rate at high
redshifts (Hughes et al.\ 1998).

If we assume instead that the bulk of the star formation traced by the
submm population is completely missed in the UV, we must add their
contribution to UV-based estimates to obtain the total star-formation
density.  Unfortunately, we still lack detailed information about the
properties of the majority of submm galaxies (most crucially redshifts)
and for this reason we therefore follow the approach of Blain et
al.\ (1999b, 1999c) and use two simple evolutionary models to derive
the history of star-formation activity using all available far-IR and
submm background and count data. First, a model of pure-luminosity
evolution of the low-redshift 60-$\mu$m {\it IRAS} luminosity function
as described in detail by Blain et al.\ (1999b); second, a model of
galaxy formation by hierarchical clustering, in which powerful episodes
of dust-enshrouded luminosity are triggered by major mergers, as
described by Blain et al.\ (1999c). The results of both models have
been updated to include count and redshift data that was unavailable
when these works were published, as described in \S6.5, are shown in
Fig.~12.

Subject to systematic uncertainties in the IMF (\S6.5.3), and the
fraction of energy released in dust-enshrouded galaxies due to AGN
accretion as compared with high-mass star-formation activity, the total
amount of star-formation in dust-enshrouded galaxies appears to exceed
that within optical- and near-IR-selected galaxies at high redshift by
a factor of several (Fig.~12).  Even after correcting the model
results, assuming that at most 20\,per cent of the emission comes from
purely gravitationally-powered emission from AGN (\S6.4; Almaini et
al.\ 1999, 2001), the obscured activity in the far-IR selected galaxy
population will still dominate the budget of massive star formation at
$z\sim 2$--3 (Hughes et al.\ 1998).  The properties of these galaxies
are thus central to our understanding of the formation of massive stars
and metals in the high-redshift Universe.

The classes of distant galaxies selected through submm and UV
techniques clearly have very different properties and represent
relatively independent subsets of the high-redshift galaxy population,
except for rare composite systems such as SMM\,J14011+0252.  What is
the physical basis for these differences?  Do the submm galaxies
represent the most massive members of the galaxy population as a whole,
or do they represent a particular evolutionary phase (Shu et al.\ 2001;
Fardal et al.\ 2001; Granato et al.\ 2001)?

The dynamical information available for several Class-II submm galaxies
shows them to be massive systems, $\gs 10^{11}$\,M$_\odot$ within 20\,kpc. 
This is nearly an order of magnitude more massive than typical UV-selected
galaxies at $z>2$ which have a comparable volume density (Pettini et
al.\ 1998; Moorwood et al.\ 2000; Kobulnicky \& Koo 2001), although a
small proportion of equally massive LBGs are known (e.g.\ Pettini et al.\
1998, 2001).  The rest-frame optical sizes of the Class-II submm galaxies
also appear to be larger than is typically found for UV-selected sources
(compare the results in Ivison et al.\ 2000a and Dickinson 2000) with
several showing merger-like morphologies.

Thus, while there is some evidence to suggest that the submm galaxies
have more massive progenitors than the typical LBGs (see also Shu et
al.\ 2001), we are more inclined to believe that they represent a
particular phase in the evolution of massive galaxies at high
redshifts, rather than a completely separate population.

The Class-II submm galaxies are optically luminous and yet gas-rich,
indicating that they possess both a significant mass of existing stars,
and a substantial reservoir of fuel for the formation of additional
stellar populations. This suggests that the present-day descendents of
these galaxies will be some of the most massive stellar systems in the
local Universe.  It is the properties of these Class-II galaxies which
are at the heart of the argument that submm galaxies evolve into the
most luminous and massive spheroids ($\gs L^\ast$) at the present day.
The merger-like morphologies of some of these galaxies is further
evidence for their role in forming the most massive local ellipticals.
Support for this argument for the general submm population is entirely
circumstantial: their strong clustering (Ivison et al.\ 2000b; Almaini
et al.\ 2001) and equivalent space density to local luminous ellipticals
(Lilly et al.\ 1999).  However, to construct a more direct evolutionary
connection will require kinematically-selected samples of high-redshift
galaxies, and as such will be observationally demanding.

Further depletion of the gas reservoir in the Class-II galaxies and
the destruction of obscuring dust by intense radiation fields would
naturally result in them evolving into massive examples of LBGs,
QSO host galaxies or luminous radio galaxies (Dunlop 2001; Granato et
al.\ 2001).  If the Class-II sources represent the most evolved submm
galaxies, then the pressing question is the relationship between these
and the majority of the submm population, which fall into Class~0/I.
Do these optically-fainter systems represent either higher redshift
systems, or an earlier phase in the growth of the Class-II galaxies,
i.e.\ does the Class 0--II classification scheme have an evolutionary
basis akin to that for protostars, or are Class 0/I galaxies simply
their less massive cousins, or intrinsically more obscured?

Our sample shows slight differences between Class-0/I and Class-II
sources, with the suggestion that the former are typically fainter in
the submm and exhibit higher radio--submm indices,
$\alpha^{850}_{1.4}$. The simplest interpretations are that the
Class-0/I sources include some galaxies that either are more obscured,
or have intrinsically colder dust, or lie at higher redshifts than the
Class-II sources.  We note that the median redshift of the Class-II
galaxies is $<\!  z\!>=1.9\pm 0.6$, compared to the estimate of $<\!
z\!>=2.8^{+2.1}_{-1.2}$ from the radio--submm index-based estimates for
the Class-0/I sources.  This hints that at least some of the Class-0/I
galaxies may be at high redshifts, $z\gs 3$, where $K$-corrections
would dim their near-IR counterparts significantly (Fig.~8; Dey et
al.\ 1999).  However, the current sample is too small and the
uncertainties too large to clearly show any differences between the
classes or definitively state the reasons for these.

%
%
\section{Conclusions}

In summary, the  bulk of the background radiation intensity has been
resolved into discrete submm sources using SCUBA at an 850-$\mu$m
intrinsic flux limit of $\sim 1$\,mJy.  The counterparts to these submm
sources appear to be dusty, ultraluminous galaxies with very diverse
optical and near-IR properties.  The majority of these galaxies appear
to lie at $z\gs 1$, with a median redshift of $z\sim 2.5$--3.

The optically brighter sources have been studied in more detail using
mm-wave interferometers, and they show the large dynamical masses and
high gas fractions expected for young massive galaxies.  The
characteristics of these galaxies are consistent with them being the
progenitors of the most massive elliptical galaxies seen in the local
Universe.  A comparison of the detailed properties of a handful of
these galaxies with local ULIRGs suggests that the rapid increase in
dust-obscured activity at high redshifts, need to explain the submm
counts and the FIRB, has its origin in the increasing instability of
the gas-rich, bulge-weak progenitors of the submm population at high
redshifts.  However, the majority of the mJy submm population remain
elusive; they have very faint (or invisible) counterparts in the
optical and near-IR and progress in investigating their nature and
properties is likely to be slow.

In the future we look forward to increases in the number of submm
galaxies with accurate redshift determinations -- crucial for follow-up
CO line-mapping to provide dynamical masses and gas fractions.
Work in this area will require deep spectroscopy in the near-IR
(and optical) on 10-m class telescopes for the brighter Class-I and II
sources, as well as more innovative approaches, such as blind radio
searches for OH/H$_2$O maser emission (Townsend et al.\ 2001).
Improved constraints on the redshift distribution of the whole submm
population await the confirmation of the radio-submm spectral index
(and more detailed SED fitting) as a reliable estimator of redshift for
submm-selected galaxies.  In part this will rely on checking at
higher luminosities and redshifts, using {\it SIRTF} and SOFIA,  the
relatively weak dependence of dust temperature on luminosity seen in
low-redshift {\it IRAS}-selected galaxies by Dunne et al.\ (2000).

Equally essential is the detailed study of the characteristics of
individual sources.  Here the main advances are likely to come from
observations across a wide range of wavelength, with the X-ray, mid-IR
and far-IR wavebands being the most promising, providing crucial
information about the distribution of dust temperatures and the power
sources driving these systems.  Observations at higher spatial
resolution, in the near-IR, mm and radio wavebands, will allow us to
study the internal structure of these galaxies -- to search for
morphological evidence of the events which triggered their prodigous
activity.  In particular, the refurbishment of {\it NICMOS} on-board
{\it HST} will provide a powerful tool to interpret the rest-frame
optical morphologies of these galaxies and compare them to local ULIRGs
to test if the same physical processes are responsible for
ultraluminous systems at low and high redshifts.  In the longer term,
the direct study of the detailed astrophysics of submm galaxies will
benefit immensely from the sensitivity and resolving power of the
10-milliarcsec resolution ALMA interferometer array.

We expect that submm surveys which exploit lens amplification, such as
the one presented here, will retain a central role in studying the
submm galaxy population.  In part this is because the properties of
these galaxies tax the capabilities of current instrumentation in many
wavebands and hence the boost provided by the lens is essential for
successful follow-up. Moreover, observations through massive
gravitational lenses allow us to probe intrinsically fainter submm
sources, which are more representative of the population responsible
for the bulk of the FIRB.  We look forward to continued exploitation of
the sample presented here, and the results of new surveys, to study the
nature of the faint submm population and answer some of the questions
raised in this paper.

Finally, the goal of future theoretical work in this area should be to
incorporate both the obscured submm population (Class~0 and I) and the
less-obscured systems (Class-II submm galaxies and the more massive
classical LBGs) into a single evolutionary sequence and hence naturally
explain the relation between the two populations.  Important
observational input on this question can be obtained by studying the
relative clustering of the various populations in well-defined
environments at high redshifts, in particular the overdense regions
around some luminous radio galaxies (Ivison et al.\ 2000b) may evolve
into the cores of massive clusters at the present day.  Such studies
will require wide-field surveys covering the UV, near-IR and longer
wavelengths, the new SCUBA2 submm camera for the JCMT and the WFCAM
panoramic near-IR camera for UKIRT will be a powerful facilities for
obtaining the essential observations in these wavebands.

\section*{Acknowledgements}

We would like to thank Amy Barger, Len Cowie, Alastair Edge, Dave Frayer,
Katherine Gunn, Frazer Owen and Ian Robson for help in undertaking this
project.  We also thank Bruno Altieri, Carlton Baugh, Chris Carilli,
Scott Chapman, Shaun Cole, Danny Dale, James Dunlop, Harald Ebeling,
Steve Eales, Richard Ellis, Andy Fabian, Carlos Frenk, Allon Jameson,
Cedric Lacey, Andy Lawrence, Simon Lilly, Malcolm Longair, Leo Metcalfe,
Chris Mihos, Glenn Morrison, Matt Page, Max Pettini, Bianca Poggianti,
Michael Rowan-Robinson, Nick Scoville, Jason Stevens, Neil Trentham, Paul
van der Werf and Min Yun for useful conversations and help.  Finally,
we acknowledge the first anonymous referee for her comments and thank the
second referee for their thorough reading of this manuscript
and suggestions for improvements.  IRS acknowledges support from the
Royal Society and the Leverhulme Trust, RJI from PPARC, AWB thanks the
Raymond and Beverly Sackler Foundations and JPK the CNRS.

%
%
{\small
\setcounter{table}{3}
\begin{table*}
\begin{center}
\caption{Photometry and limits ($2\sigma$)}
\begin{tabular}{lccccccl}
\noalign{\smallskip}\hline
\noalign{\smallskip}
{Source} & {$S_{850}$} &  {T$_{\rm UKIRT}$} & {$K$} & {$(R-K)$} & {$(I-K)$} & Amp. & { Comments ~\hfill } \cr
{} & {(mJy)} & {(ks)} & {}  & {} & {} & {} & {} \cr
\noalign{\smallskip}\hline
\noalign{\smallskip}
SMM\,J02399$-$0136 & 23.0  & ~4.3 &  17.8$\pm$0.1 & 3.4$\pm$0.1 & 2.7$\pm$0.1 & ~2.5 & L1/L2$^a$ \cr
SMM\,J00266+1708   & 18.6  & ~3.2 &  ~22.5$\pm$0.1$^{b}$ & ...   & $>3.7$      & ~2.4 & M12$^{b}$ \cr
SMM\,J09429+4658   & 17.2  & ~8.1 &  19.4$\pm$0.3 & $>4.4$      & $>6.0$     & ~2.0 & H5$^{c}$ \cr
SMM\,J14009+0252   & 14.5  & ~9.2 &  21.0$\pm$0.4 & $>5.6$      & $>2.2$      & ~1.5 & J5$^{d}$, $R$ limit from new {\it WFPC2} data \cr
SMM\,J14011+0252   & 12.3  & ~7.0 &  17.6$\pm$0.1 & 3.7$\pm$0.1 & 3.1$\pm$0.1 & ~2.8 & J1/J2$^{d}$  \cr
SMM\,J02399$-$0134 & 11.0  & ~9.2 &  16.3$\pm$0.2 & 4.3$\pm$0.2 & 3.0$\pm$0.3 & ~2.5 & L3$^{e}$ \cr
SMM\,J22471$-$0206 & ~9.2  & 19.3 &  17.5$\pm$0.1 & ...         & 4.0$\pm$0.1 & ~1.9 & P4? \cr
SMM\,J02400$-$0134 & ~7.6  & ~9.2 &  $>21.6$      & ...         & ...         & $>1.9$~ & Blank$^f$, $I>26$    \cr
SMM\,J04431+0210   & ~7.2  & ~3.2 &  19.1$\pm$0.1 & $>6.7$      & $>6.7$      & ~4.4 & N4$^c$ \cr
SMM\,J21536+1742   & ~6.7  & 10.4 &  $>21.6$      & ...         & $<3.8$      & ~1.9 & K2? $I\sim 25.4$  \cr
SMM\,J00265+1710   & ~6.1  & ~4.8 &  $>21.0$      & ...         & ...         & $>2.3$~ & Blank?, $I>25$  \cr
SMM\,J22472$-$0206 & ~6.1  & ~7.6 &  $>21.4$      & ...         & $<2.4$      & ~2.2 & P2? $I=23.8$ \cr
SMM\,J00266+1710   & ~5.9  & ~3.2 &  $>20.9$      & ...         & ...         & $>3.6$~ & Blank?, $I>26$     \cr
SMM\,J00267+1709   & ~5.0  & ~3.2 &  $>20.7$      & ...         & ...         & $>2.2$~ & Blank$^f$, $I\gs 25.5$ \cr
SMM\,J04433+0210   & ~4.5  & ~3.2 &  $>20.8$      & ...         & $<3.4$      & ~1.5 & N5 $I=24.2$\cr
\noalign{\medskip}
\multispan8{Cluster Galaxies \hfil} \cr
\noalign{\smallskip}
SMM\,J21536+1741   & ~9.1  & 10.4 &  13.2$\pm$0.0 & ...         & 2.5$\pm$0.0 & ... & A\,2390 cD$^{g}$ \cr
SMM\,J14010+0252   & ~5.4  & ~7.0 &  12.4$\pm$0.0 & ...         & 3.0$\pm$0.0 & ... & A\,1835 cD$^{g}$ \cr
\noalign{\smallskip}\hline
\noalign{\smallskip}
\end{tabular}
\end{center}
$a$) Ivison et al.\ (1998a) --
$b$) Frayer et al.\ (2000) --
$c$) Smail et al.\ (1999a) --
$d$) Ivison et al.\ (2000a) --
$e$) Soucail et al.\ (1999) -- \\
$f$) Smail et al.\ (1998) --
$g$) Edge et al.\ (1999).
\end{table*}
}

%
%
{\small
\setcounter{table}{4}
\begin{table*}
\begin{center}
\caption{Spectroscopy and redshift constraints}
\begin{tabular}{lccrccccl}
\noalign{\smallskip}\hline
\noalign{\smallskip}
{Source} & {$S_{850}$} & {$S_{1.4}$} & {$\alpha_{1.4}^{850}$~~~~} & {$z_\alpha$} & \multispan2{~$z_{\rm spec}^e$} & {$z_{\rm SED}$} & {Comments ~\hfill } \cr
{} & {(mJy)} & {($\mu$Jy)}  & {} & {} & {Reliable} & {Possible} & {} & {} \cr
\noalign{\smallskip}\hline
\noalign{\smallskip}
SMM\,J02399$-$0136 & 23.0  & 526    & $0.70\pm 0.02$  & 0.9--2.3 & 2.80 & ... &  ... & L1/L2 -- Sy 2 merger$^a$ \cr
SMM\,J00266+1708   & 18.6  & 100    & $0.95\pm 0.04$  & 2.0--5.0 & ... & ... &  2--5 & M12 -- ERO?$^{b}$ \cr
SMM\,J09429+4658   & 17.2  & ~32    & $1.14\pm 0.07$  & $>$3.4   & ... & ... &  2--4 & H5 -- ERO$^{c}$ \cr
SMM\,J14009+0252   & 14.5  & 529    & $0.60\pm 0.03$  & 0.7--1.8 & ... & ... &  3--5 & J5$^c$ -- ERO$^{d}$ \cr
SMM\,J14011+0252   & 12.3  & 115    & $0.85\pm 0.05$  & 1.5--3.6 & 2.56 & ... &  ... & J1/J2 -- starburst merger$^{d}$  \cr
SMM\,J02399$-$0134 & 11.0  & $\sim$500 & $0.55\pm 0.04$  & 0.6--1.5 & 1.06 & ... & ... & L3 -- Sy 1.5/2 ring galaxy$^{f}$ \cr
SMM\,J22471$-$0206 & ~9.2  & $<$65  & $>0.90\pm 0.08$   & $>$1.7 & ... & 1.16 & ... & P4? -- weak AGN? \cr
SMM\,J02400$-$0134 & ~7.6  & $<$33 & $>0.99\pm 0.11$  & $>$2.2 & ... & ... &  ... & Blank field$^g$    \cr
SMM\,J04431+0210   & ~7.2  & $<$70  & $>0.84\pm 0.10$  & $>$1.4 & ... & ... & 2--4 & N4 -- ERO$^c$ \cr
SMM\,J21536+1742   & ~6.7  & ... & ...~~~~~~  & ... & ... & ~1.60? & ... & K2?  \cr
SMM\,J00265+1710   & ~6.1  & $\leq$110   & $\geq 0.73\pm 0.07$  & $>$1.0 & ... & ... & ... & Blank field?  \cr
SMM\,J22472$-$0206 & ~6.1  & $<$50 & $>0.87\pm 0.12$  & $>$1.6 & ... & ~2.11? &  ... & P2? \cr
SMM\,J00266+1710   & ~5.9  & $<$33 & $>0.93\pm 0.12$  & $>$1.9 & ... &  ... & ... & Blank field?     \cr
SMM\,J00267+1709   & ~5.0  & $<$30 & $>0.92\pm 0.13$  & $>$1.8 &  ... & ... & ... & Blank field$^g$ \cr
SMM\,J04433+0210   & ~4.5  & ~70     & $0.75\pm 0.12$  & 1.1--2.7 & ... & ... &  ... & N5 \cr
\noalign{\medskip}
\multispan7{Cluster Galaxies \hfil} \cr
\noalign{\smallskip}
SMM\,J21536+1741   & ~9.1  & $226 \times 10^3$ & $-0.58\pm 0.05$  & ... & 0.23 & ... & ... & A\,2390 cD$^{h}$ \cr
SMM\,J14010+0252   & ~5.4  & $28.6\times 10^3$  & $-0.30\pm 0.06$ & ... & 0.25 & ... & ... & A\,1835 cD$^{h}$ \cr
\noalign{\smallskip}\hline
\noalign{\smallskip}
\end{tabular}
\end{center}
$a$) Ivison et al.\ (1998a) --
$b$) Frayer et al.\ (2000) --
$c$) Smail et al.\ (1999a) --
$d$) Ivison et al.\ (2000a) --
$e$) Barger et al.\ (1999a) -- \\
$f$) Soucail et al.\ (1999) --
$g$) Smail et al.\ (1998) --
$h$) Edge et al.\ (1999).
\end{table*}
}

\end{document}